\documentclass[11pt]{article}

\usepackage[margin=1in]{geometry}
\usepackage[T1]{fontenc}
\usepackage[utf8]{inputenc}
\usepackage{lmodern}
\usepackage{microtype}

\usepackage{amsmath,amssymb}
\usepackage{graphicx}
\usepackage{siunitx} 

\usepackage[numbers,sort&compress]{natbib}
\usepackage{hyperref}   
\usepackage{cleveref}   

\usepackage{authblk}

\usepackage{adjustbox}
\usepackage{algorithm}
\usepackage{algpseudocode}
\usepackage{amsfonts}
\usepackage{amsmath}
\usepackage{amssymb} 
\usepackage{amsthm}
\usepackage[greek,english]{babel}
\usepackage{booktabs}
\usepackage{cleveref}
\usepackage{colortbl}
\usepackage{comment}
\usepackage{dutchcal}
\usepackage{enumerate}
\usepackage{epsfig}
\usepackage{graphicx}
\usepackage{hyperref}
\usepackage[utf8]{inputenc}
\usepackage{mathtools} 
\usepackage[framemethod=TikZ]{mdframed}
\usepackage{multirow}
\usepackage{siunitx} 
\usepackage{tikz}
\usepackage{url}
\PassOptionsToPackage{table}{xcolor}
\usepackage{xfrac} 
\usepackage{xspace}
\usetikzlibrary{bayesnet}
\usetikzlibrary{positioning}

\usepackage[show]{chato-notes}

\graphicspath{{figures/}}

\newcommand{\spara}[1]{\smallskip\noindent{\bf #1}}

\theoremstyle{thmstyleone}%
\theoremstyle{thmstyletwo}%
\theoremstyle{thmstylethree}%

\algnewcommand\algorithmicforeach{\textbf{for each}}
\algdef{S}[FOR]{ForEach}[1]{\algorithmicforeach\ #1\ \algorithmicdo}

\renewcommand{\vec}[1]{\mathbf{#1}}

\definecolor{gold}{rgb}{1.0, 0.84, 0.0}  
\definecolor{silver}{rgb}{0.75, 0.75, 0.75}  
\definecolor{bronze}{rgb}{0.8, 0.5, 0.2}  

\title{Uncovering the Sociodemographic Fabric of Reddit}

\author[1]{Federico Cinus\thanks{Correspondence: \href{mailto:federico.cinus@centai.eu}{federico.cinus@centai.eu}}}
\author[1]{Corrado Monti}
\author[1]{Paolo Bajardi}
\author[1]{Gianmarco De Francisci Morales}

\affil[1]{CENTAI, Corso Inghilterra 3, 10138 Turin, Italy}

\date{} 

\begin{document}
\maketitle

\begin{abstract}
Understanding the sociodemographic composition of online platforms is essential for accurately interpreting digital behavior and its societal implications.
Yet, current methods often lack the transparency and reliability required, risking misrepresenting social identities and distorting our understanding of digital society.
Here, we introduce a principled framework for sociodemographic inference on Reddit that leverages over \num{850000} user self-declarations of age, gender, and partisan affiliation.
By training models on sparse user activity signals from this extensive, self-disclosed dataset, we demonstrate that simple probabilistic models, such as Naive Bayes, outperform more complex embedding-based alternatives.
Our approach improves classification performance over the state of the art by up to 19\% in ROC AUC and maintains quantification error below 15\%.
The models produce well-calibrated and interpretable outputs, enabling uncertainty estimation and subreddit-level feature importance analysis.
More broadly, this work advocates for a shift toward more ethical and transparent computational social science by grounding sociodemographic analysis in user-provided data rather than researcher assumptions.
\end{abstract}


\section{Significance statement}
Understanding the sociodemographic composition of online platforms is central to computational social science.
Yet for Reddit, one of the world's largest discussion platforms, transparent and reliable demographic inference tools have been lacking.
We address this gap by collecting large-scale self-declared data and extensively training and testing models to infer age, gender, and political leaning.
Our Bayesian framework produces accurate, interpretable predictions and supports both individual- and population-level analysis, including calibrated uncertainty and privacy-conscious quantification strategies.
Designed for computational social scientists, digital demographers, and platform researchers, this work provides a scalable and ethically grounded foundation for analyzing social patterns in digital communities.

\section{Introduction: Inferring Online Socio-Demographics}
\label{sec:intro}

\begin{figure*}[t]
    \centering
    \begin{minipage}[t]{0.32\textwidth}
        \centering
        \includegraphics[width=\linewidth]{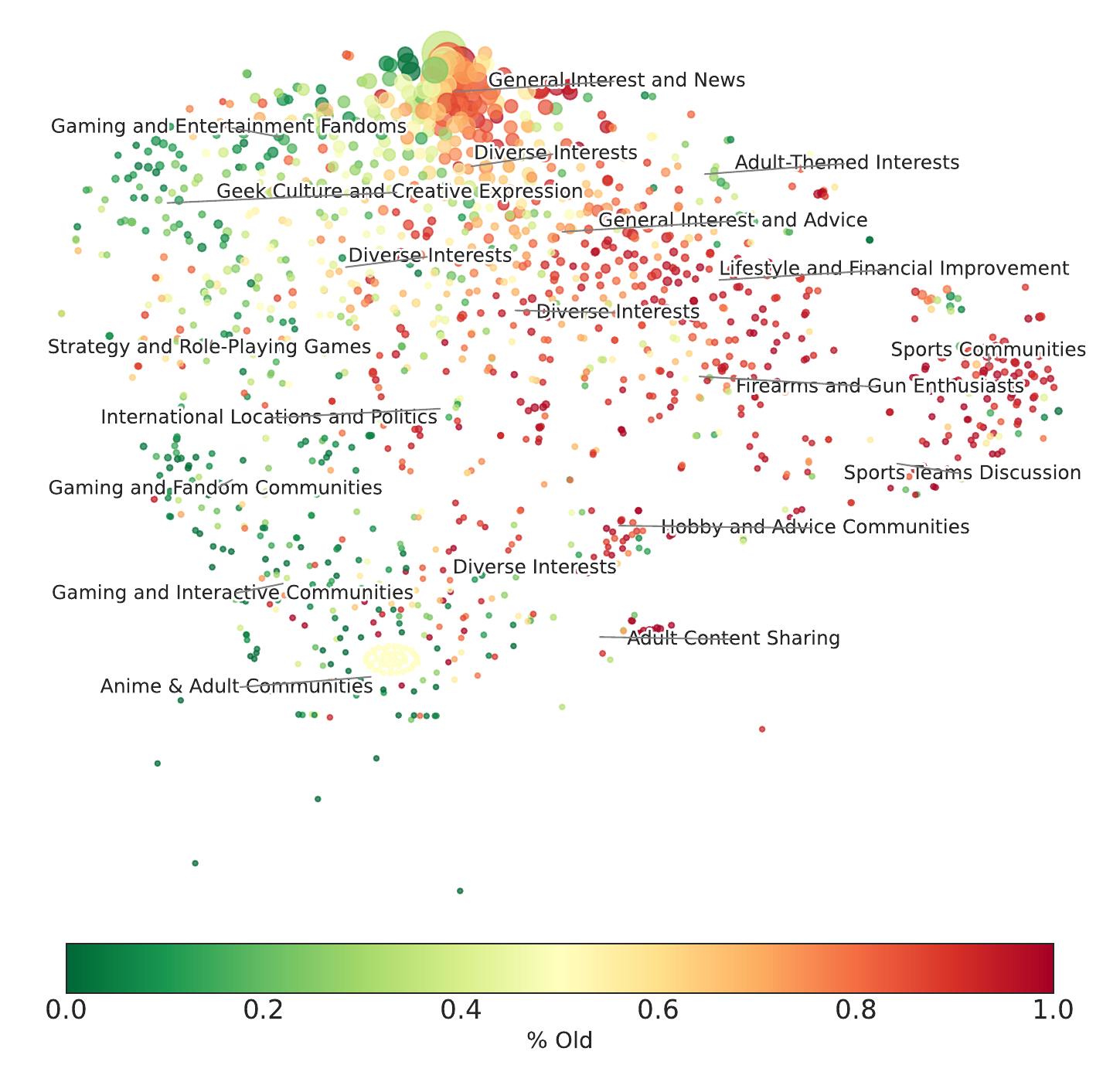}
        \textbf{(a)} Age
    \end{minipage}
    \hfill
    \begin{minipage}[t]{0.32\textwidth}
        \centering
        \includegraphics[width=\linewidth]{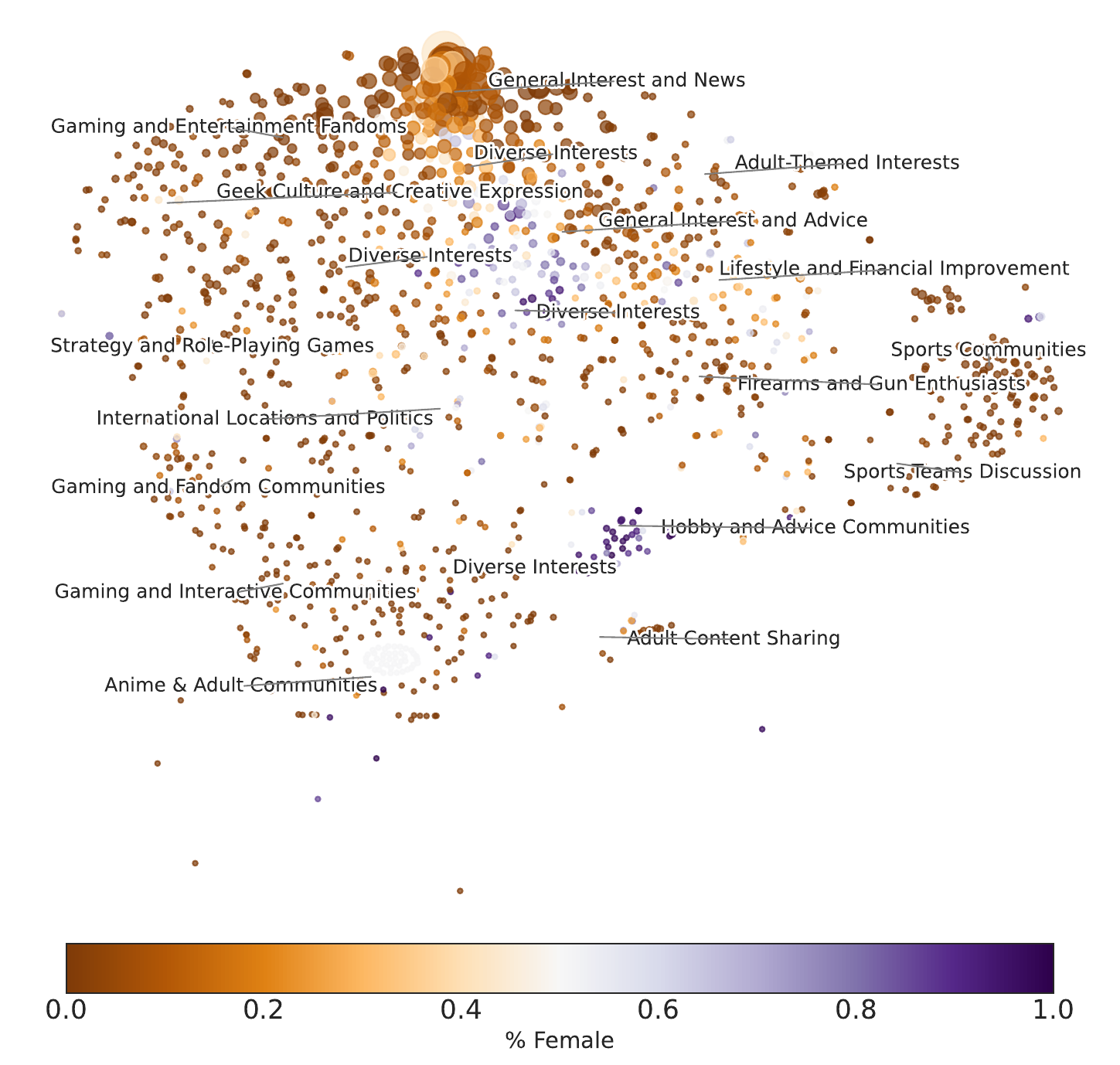}
        \textbf{(b)} Gender
    \end{minipage}
    \hfill
    \begin{minipage}[t]{0.32\textwidth}
        \centering
        \includegraphics[width=\linewidth]{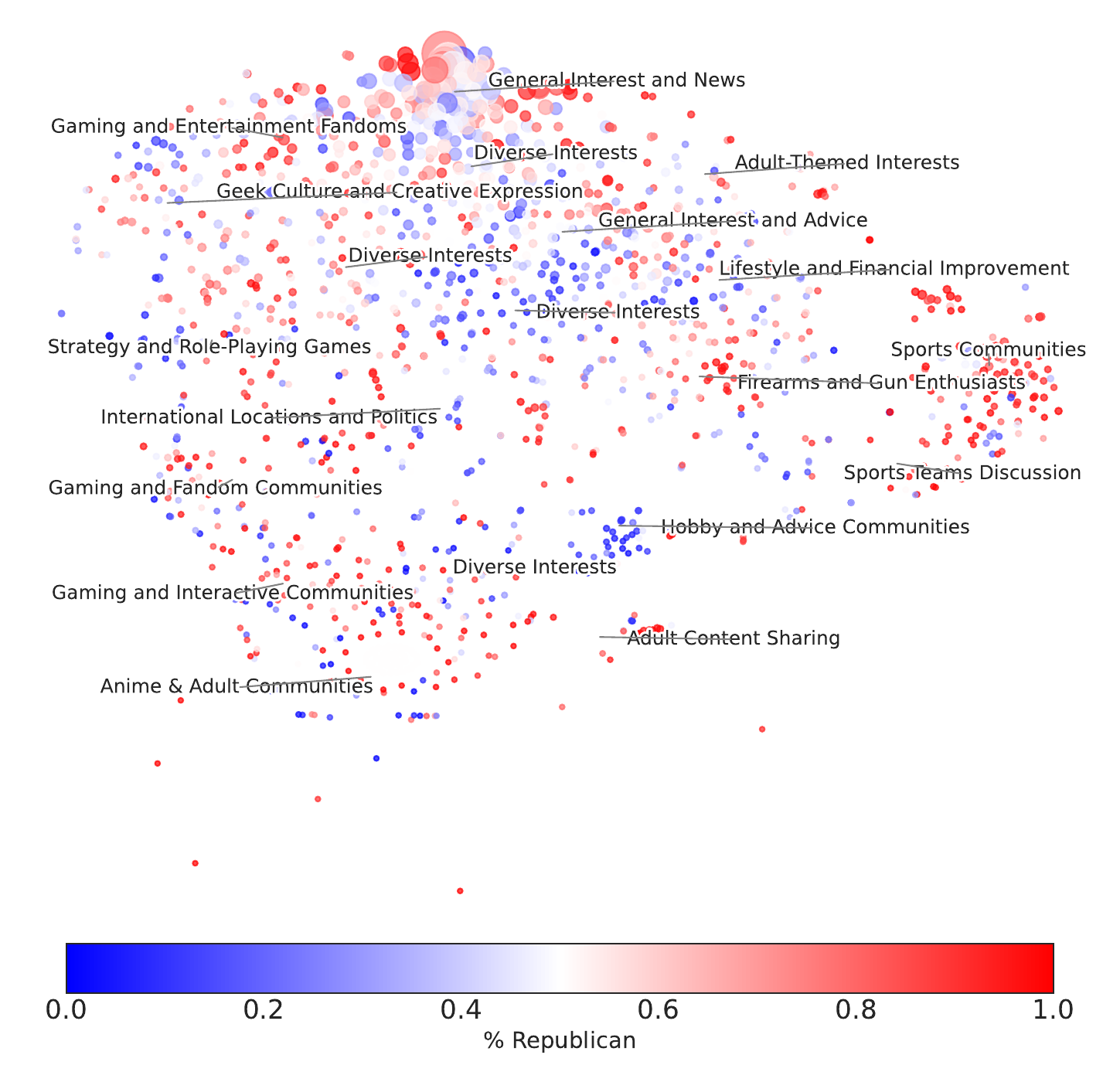}
        \textbf{(c)} Partisan Affiliation
    \end{minipage}
    \caption{\textbf{Sociodemographic Fabric of Reddit: 2D projection of subreddits colored by sodemographic attributes.} We visualize a network of co-participation patterns in 2019 Reddit data among 1,400 subreddits, chosen by popularity. Posts and comments were sampled at 5\% and filtered to exclude deleted content and inactive users. We retained users with at least 100 contributions and classified them by age, gender, and partisanship using pre-trained Naive Bayes models with Classify \& Count quantifier. Subreddit positions are computed via t-SNE on a PPMI-transformed co-occurrence matrix. Colors indicate the share of users in each subreddit predicted to be (a) older than the median age, (b) female, or (c) Republican-leaning. Marker size reflects subreddit activity, and labels represent descriptive summaries of the subreddits in each of 20 KMeans clusters, generated using GPT-4o. An interactive version of the plot is available at \url{https://federicocinus.github.io/reddit-fabric}.}
    \label{fig:fabric}
\end{figure*}

A core challenge in social media research is understanding the connection between the online and offline behaviors of individuals.
This connection is critical to assessing how digital environments shape real-world outcomes---such as the influence of information diets on voting behavior~\citep{bach2021predicting}, or the impact of algorithmic curation on vaccine hesitancy~\citep{sasse2024understanding}.
Sociodemographic attributes serve as a key bridge between these two realms, linking online activity to offline identities and social contexts.

Individuals with similar sociodemographic backgrounds often share environments, routines, and concerns~\citep{fan2023diversity}.
These shared experiences shape not only how people perceive the world~\citep{lee2014newsworthy} but also how they form opinions~\citep{hamilton2011all, capozzi2021clandestino}.
As such, sociodemographic attributes are widely used to contextualize user behavior on digital platforms~\citep{rivas2020classification, gjurkovic2021pandora, tadesse2019detection}, and to understand broader dynamics of polarization, participation, and influence~\citep{monti2023evidence, colacrai2024navigating}.

Beyond their descriptive utility, sociodemographic attributes serve a methodological function.
They help adjust for hidden confounders in causal analysis, and they enable post-stratification to align online observations with population-level trends~\citep{giorgi2022correcting}.
From a normative perspective, sociodemographic analysis can also promote inclusion by surfacing underrepresented voices and correcting for structural imbalances in visibility and participation of minorities.

\smallskip
In this context, sociodemographic inference (SDI)---the recovery of latent sociodemographic attributes---has become a foundational task in computational social science (CSS).
Yet not all methods are equally suited for scientific analysis.
To be effective, SDI approaches must be \emph{reliable} (producing few mistakes and consistent outputs), \emph{transparent} (supporting interpretability and explainability for human understanding and auditing), and \emph{efficient} (scaling to large social datasets with reasonable computational cost).
We define \emph{interpretability} as the presence of calibrated output scores that allow direct understanding of a model's predictions, and \emph{explainability} as access to internal model parameters---such as feature weights---that support downstream analysis and auditing of the relationship between features and predictions.

This work addresses two core SDI tasks based on user activity patterns.
The first, \emph{classification}, aims to predict individual-level sociodemographic attributes---such as gender, year of birth, or political affiliation.
The second, \emph{quantification}, estimates the aggregate prevalence of those attributes within a given user cohort.
While classification supports fine-grained modeling and stratified analysis, quantification enables population-level insights while preserving individual privacy. 
This capability is especially important in public health, policy, and ethics-driven domains, where de-identification and aggregation are critical for responsible use.

More broadly, sociodemographic inference benefits from a holistic approach---integrating individual-level prediction with aggregate estimation, and linking model outputs to social interpretation. This perspective shifts focus from narrow accuracy metrics to interpretability, ethical grounding, and the societal implications of modeling identity online.

\subsection{Reddit}  
Among major platforms, Reddit presents a compelling setting for SDI.
Self-described as ``The front page of the Internet,'' Reddit ranks among the top global websites and functions as a central hub for information gathering and discussion.
Its content is organized into \emph{subreddits}, thematic communities that structure participation around shared interests, identities, and norms~\citep{baumgartner2020pushshift}.
User participation in subreddits provides a clear, interpretable signal for modeling, offering a scalable and semantically meaningful input for SDI tasks~\citep{medvedev2019anatomy}.

Reddit's public data and diverse communities have made it a rich resource for academic inquiry.
It has been used to study public health~\citep{balsamo2019firsthand, lokala2022computational, balsamo2023pursuit}, deliberation and ideological dynamics~\citep{cinelli2021echo, monti2022language, colacrai2024navigating}, and extremist or conspiratorial discourse~\citep{klein2019pathways, rollo2022communities}.
Yet, despite this attention, little systematic work has evaluated how demographic attributes can be reliably and transparently inferred from Reddit activity.
While some studies rely on demographic proxies, few examine whether the methods themselves are interpretable, auditable, and scalable.

\subsection{Literature Gap}  
Existing approaches to sociodemographic inference (SDI) on social media typically fall into two camps: highly attribute-specific models and general-purpose embedding-based methods.  
The former are often limited to a single attribute~\citep{wang2018s,chew2021predicting,vasilev2018inferring} and rely on domain-specific features or assumptions, such as linguistic cues from user text. While these models can perform well in narrow contexts, they are difficult to generalize and typically do not scale across platforms or demographic categories.

At the other end of the spectrum, general-purpose frameworks based on user activity have been proposed---most notably by~\citet{waller2021quantifying}.
Their method constructs an embedding space from subreddit co-participation patterns and then projects it onto sociodemographic axes defined via manually selected ``seed'' subreddits. 
This approach has gained widespread adoption~\citep{colacrai2024navigating, monti2023evidence, mok2023echo, kim2025capturing} and provides broad coverage, but has several limitations.
It is fully unsupervised, and its validation is largely indirect or circumstantial.
Its dependence on arbitrarily chosen seed subreddits may introduce bias and reduce generalizability.
Additionally, the use of neural embeddings (e.g., Word2Vec) renders the model opaque and difficult to audit or interpret---an issue for applications requiring transparency and accountability.

While some prior work leverages user-generated text~\citep{wang2018s,chew2021predicting,vasilev2018inferring}, our approach focuses instead on sparse participation signals: which subreddits a user engages with. 
This choice offers multiple advantages: it is more scalable, less invasive, and easier to interpret, making it suitable for population-level inference.

Ultimately, this line of work underscores a central tension in SDI research: the trade-off between \emph{generalizability}, \emph{reliability}, and \emph{efficiency}.
Few existing methods achieve all of them, especially on Reddit, where the complexity and volume of user behavior demand approaches that are not only scalable but also transparent and empirically grounded.

\subsection{Summary of Data and Methods}
To address these limitations, we introduce a framework for sociodemographic inference (SDI) that is transparent, scalable, and grounded in Reddit users' own self-declarations.

We collect a large-scale dataset of public self-declared demographic statements from user comments, covering age, gender, and political affiliation.
These disclosures form a large and verifiable dataset that reflects Reddit's user base without relying on externally-imposed categories.

Using this dataset, we evaluate a broad set of models for both individual-level prediction (classification) and group-level estimation (quantification).
Our benchmarks include simple probabilistic models, decision-tree-based approaches, and widely used embedding-based baselines.
We assess model performance using standard accuracy metrics and place particular emphasis on interpretability: examining how model outputs and internal signals can support auditing, explanation, and uncertainty estimation.

Overall, we find that simple, interpretable models not only exceed the performance of more complex alternatives but also offer greater transparency, better uncertainty estimation, and higher robustness in data-sparse settings.

\subsection{Contributions}  
This work promotes both a methodological contribution and an epistemological shift.

Methodologically, we introduce a framework for sociodemographic inference that is \emph{reliable} and \emph{efficient}.
We show that simple supervised models trained on self-disclosed data outperform embedding-based baselines---even under distant supervision.
Our Multinomial Naive Bayes model achieves ROC AUC gains of 17\%, 19\%, and 6\% for age, gender, and partisan classification, respectively.
In quantification tasks, it maintains mean absolute errors below 15\% on large-scale data and under 18\% in sparse conditions.
The model remains robust with as few as \num{1000} labeled instances, and we find that increasing model complexity or adding unsupervised co-participation data yields little additional benefit.
These results underscore the efficiency and scalability of interpretable probabilistic methods.

Epistemologically, we argue that demographic categories should not be predefined and imposed, but instead allowed to emerge from the community itself.
Self-disclosures offer a bottom-up alternative to biased or arbitrary class definitions. Our framework foregrounds \emph{transparency} through \emph{interpretability} (via calibrated outputs), \emph{explainability} (via feature importance), and \emph{auditability} (through clearly defined model assumptions and mechanisms).

Finally, we demonstrate the utility of our framework by reproducing Reddit's sociodemographic `fabric' across \num{1400} active subreddits (\Cref{fig:fabric}).
Using predicted user attributes, we recover coherent and interpretable demographic patterns: sports communities tend to skew older and more male; gaming subreddits are younger and male; hobby and advice communities are more female and left-leaning; firearms-related subreddits are predominantly male and right-leaning.
These patterns align with intuitive social divisions, demonstrating a scalable and interpretable approach to quantifying how identity shapes participation in social media.

\section{Background}
\label{sec:literature}
\subsection{Modeling Approaches}
Sociodemographic inference can be approached through a range of modeling strategies, each with distinct trade-offs depending on the input data and analytical goals.

In this study, we focus on models that take as input users' \emph{sparse participation signals}---specifically, counts of subreddit activity.
This feature choice offers strong scalability, preserves user privacy, and avoids the complexities and biases associated with language processing.

Probabilistic models such as Naive Bayes excel in this setting: they are computationally efficient, naturally interpretable, and handle uncertainty well.
Logistic regression, another common method, offers a simple and transparent structure, though it may struggle with multicollinearity when subreddits with overlapping user bases are used as features.
Tree-based models such as random forests can capture complex nonlinear relationships among participation patterns but often sacrifice interpretability in the process.

In contrast, research in natural language processing (NLP) has developed models that infer demographics from user-generated \emph{text}~\citep{wang2018s,chew2021predicting,vasilev2018inferring}.
These approaches can capture subtle linguistic cues and have demonstrated strong predictive accuracy, particularly when tailored to specific sociodemographic attributes.
However, they face several limitations: high computational cost (both in time and memory), limited scalability to large social media platforms, and a typical focus on a single attribute at a time.
Moreover, they raise ethical concerns around fairness, privacy, and the potential for bias in language-based representations.
Given our focus on scalable, interpretable, and multi-attribute inference from participation data, we treat NLP-based approaches as outside the scope of this study.

Finally, the current state of the art on Reddit, as introduced by~\citet{waller2021quantifying}, relies on an unsupervised embedding approach built from subreddit co-participation.
While powerful and widely adopted~\citep{colacrai2024navigating,del2023mental,monti2023evidence}, this method is opaque and difficult to audit.
It projects users into a latent space and infers demographic traits by comparing their position to hand-selected ``seed'' subreddits. This introduces potential bias and lacks clear interpretability.
Moreover, to address possible misclassifications, many studies using this approach apply only to users with extreme output scores, excluding a large portion of the population and limiting generalizability.

Conversely, our approach emphasizes models that are transparent and auditable, thus enabling both classification and quantification in an efficient and empirically-grounded way.

\subsection{General Training Strategies}  
A range of general machine learning strategies have been developed to improve model accuracy and scalability, and can be applied in the context of sociodemographic inference. To address challenges such as data sparsity and noisy labels, both supervised and semi-supervised learning approaches can be used. 
Supervised methods leverage labeled data to train predictive models, while semi-supervised techniques additionally use the abundant unlabeled data available, potentially improving generalization even with limited labeled samples.

Data augmentation techniques—such as oversampling and undersampling—are also widely used to enhance model performance~\citep{chawla2002smote,he2008adasyn}.
These methods enrich the training dataset by generating synthetic examples or balancing the dataset to reduce class imbalances and improving the model's ability to learn diverse patterns~\citep{shorten2019survey}.
Additionally, stratification can be employed to ensure that the sociodemographic distribution within the dataset is accurately represented~\citep{shahrokh2014effect}.
This step is especially critical for sensitive labels, as it helps minimize biases and ensures equitable performance across different sociodemographic groups.

\subsection{Data Foundations: Features and Labels}
Prior work on SDI has often relied on linguistic features~\citep{wang2018s,chew2021predicting,vasilev2018inferring}.
While text-based signals can offer rich information, they are computationally expensive and dense, making them difficult to scale and interpret, especially on large platforms such as Reddit.
Several recent approaches to SDI have leveraged user participation data---most notably subreddit co-participation---as predictive signals.
An influential example is \citet{waller2021quantifying}, who construct user embeddings from subreddit activity and project them onto demographic dimensions.
This line of work avoids the computational demands of text-based inference while offering broad coverage across Reddit's user base.
The main intuition is to start from participation patterns---particularly subreddit activity counts---which are sparse, transparent, and nearly universal.
They are memory-efficient, easy to extract, and support scalable matrix-based computation. 

Ground truth labels are another challenge in SDI research. Traditional surveys are reliable but costly and limited in scale. Crowdsourcing approaches~\citep{zhang2016learning,kittur2008crowdsourcing,stritch2017opportunities} offer broader reach but often suffer from annotator inconsistency and bias~\citep{eickhoff2018cognitive,hettiachchi2021investigating}, and raise ethical concerns over transparency and consent~\citep{barbosa2019rehumanized}.

\subsection{Learning Tasks}
The usage of sociodemographic inference can be broadly distinguished into two downstream tasks: classification and quantification.
Classification aims to predict user-level sociodemographic attributes based on features derived from user behavior or content.
Typically, the per-user label is then combined with some other quantity---for instance, some property of their comments~\citep{de2022social, betti2025moral}.
Common attributes of interest include age, gender, affluence, and partisan affiliation.
While effective in identifying specific user characteristics, classification methods might raise ethical concerns if such data is shared with third parties, due to their sensitive nature and potential for compromising user privacy.

In contrast, quantification focuses on estimating the prevalence of sociodemographic groups at the population level, thus offering a complementary perspective~\citep{gonzalez2017review}.
By emphasizing aggregate trends rather than individual predictions, quantification mitigates biases caused by misclassification errors in classification models and aligns more closely with the broader objectives of CSS~\citep{nature2021digital}.
For many applications, understanding group-level trends is sufficient, making quantification an essential and often more practical approach for sociodemographic research.
The distinction between classification and quantification is crucial, as the latter supports privacy-preserving analyses while maintaining the integrity of population-level insights.


\section{Sociodemographics Can Be Discovered Bottom-Up from Self-Declarations}
\begin{table*}[t]
\caption{Statistics for self-declarations, with bots and non-coherent users removed. Only subreddits matched by the attribute-specific regular expressions are included.}
\centering
\resizebox{.85\textwidth}{!}{
\sisetup{table-text-alignment=center,table-format=6.0}
\begin{tabular}{lSSSSrrS}
\toprule
{\textbf{Attribute}} & {\textbf{Declarations}} & {\textbf{Users}} & {\textbf{Inactive Users}} & {\textbf{Ambiguous Users}} & \multicolumn{2}{c}{\textbf{Class Proportion}} & {\textbf{Subreddits}} \\
\midrule
Year of Birth & 420803 & 401390 & 1630 & 17341 & Young: 43.81\% & Old: 56.19\% & 9806 \\
Gender & 424330 & 403428 & 1634 & 18337 & Male: 50.89\% & Female: 49.11\% & 9809 \\
Partisan Affiliation & 6369 & 6118 & 4 & 251 & Dem.: 54.55\% & Rep.: 45.45\% & 9137 \\
\bottomrule
\end{tabular}
}
\label{table:stats}
\end{table*}

We use subreddit participation as the primary input signal for sociodemographic inference.
To ensure robustness and generalizability, we select the \num{10000} most active subreddits based on comment volume from 2016 to 2020, consistent with prior work~\citep{waller2021quantifying}. 
This choice strikes a balance between feature richness and computational efficiency, while introducing potential multicollinearity, an issue we address during modeling.

\subsection{Bottom-Up Labeling via Self-Declarations}
To construct ground-truth labels, we extract \emph{self-declared} sociodemographic attributes directly from users' comments and submissions.
These disclosures provide high-quality, verifiable labels that reflect user intent and avoid reliance on external inference or proxy assumptions.
We refer to this property as \emph{authenticity}: transparent basis for model training.

We design targeted regular expressions to identify three attributes: age (converted to birth year), gender, and political affiliation.
For political identity, we search for explicit phrases such as ``I am a Democrat'' or ``I'm a registered Republican,'' excluding negations.
Patterns for age and gender (e.g., ``F27,'' ``20M'') are matched only when adjacent to first-person pronouns, following~\citet{de2022social}.
An overview of regex patterns is shown in \Cref{fig:regex-pol}.

Bot accounts are removed using a curated list~\cite{rollo2022communities}. Users with inconsistent declarations (e.g., conflicting gender statements over time) are also excluded---approximately 4\% of cases.
For all remaining users, we retrieve their complete Reddit activity across the selected subreddits between 2016 and 2020.

\subsection{Dataset Overview}
\Cref{table:stats} summarizes the resulting dataset.
We obtain roughly \num{400000} self-declarations each for gender and birth year, and a smaller but still substantive set for partisan affiliation.
Class distributions are relatively balanced, and the final number of subreddits with activity from labeled users remains close to the original \num{10000}.
Additional distributional statistics are presented in Materials and Methods, along with a comparison to a control group, showing that self-declaring users exhibit similar participation behavior to the broader Reddit population.

\section{Supervision from Self-Declarations Outperforms Embedding-Based Approaches}
\begin{table}[t]
\caption{
Classification results for three sociodemographic attributes (Year of Birth, Gender, Partisan Affiliation) under true supervision, reporting mean $\pm$ standard deviation over 100 bootstrap samples. We benchmark four models: 
\textbf{(i)} Majority class baseline; 
\textbf{(ii)} WA model~\citep{waller2021quantifying}, which embeds subreddit participation into demographic axes; 
\textbf{(iii)} Random Forest (RF) with 50 estimators and maximum depth of 10; and 
\textbf{(iv)} Multinomial Naive Bayes (NB) with additive smoothing ($\alpha=1$). 
Best scores in bold.
}
\centering
\resizebox{\columnwidth}{!}{
\sisetup{table-format=1.4(1),table-text-alignment=center,round-mode=uncertainty,round-precision=2,uncertainty-mode=separate,retain-zero-uncertainty,detect-weight}
\newrobustcmd{\B}{\bfseries}

\small
\begin{tabular}{ll SS}
\toprule
\text{\textbf{Attribute}} & \text{\textbf{Model}}  & \text{\textbf{F1}}  & \text{\textbf{ROC AUC}} \\
\midrule
\multirow{4}{*}{Year} & Majority     & 0.4033 \pm 0.002 & 0.4967 \pm 0.0002\\
 & NB           & \B 0.6864 \pm 0.0016 & \B0.7368 \pm 0.0017\\
 & WA           & 0.6249 \pm 0.0019 & 0.6298 \pm 0.0021\\
 & RF           & 0.5089 \pm 0.0065 & 0.7122 \pm 0.0033\\
\midrule
\multirow{4}{*}{Gender} & Majority & 0.4231 \pm 0.0022 & 0.5322 \pm 0.0007\\
 & NB       & \B 0.672 \pm 0.0018 & \B 0.7956 \pm 0.0014\\
 & WA       & 0.3752 \pm 0.0022 & 0.6667 \pm 0.0019\\
 & RF       & 0.6361 \pm 0.0037 & 0.7437 \pm 0.0038\\
\midrule
\multirow{4}{*}{Partisan} & Majority & 0.5508 \pm 0.0151 & 0.5521 \pm 0.0134\\
 & NB       & \B 0.6608 \pm 0.0132 & 0.7131 \pm 0.0145\\
 & WA       & 0.579 \pm 0.0126 & 0.6728 \pm 0.0131\\
 & RF       & 0.5626 \pm 0.027 & \B 0.7242 \pm 0.0145\\
\bottomrule
\end{tabular}

}
\label{table:classif}
\end{table}

Sociodemographic classification requires predicting individual-level demographic attributes from subreddit participation patterns.
Formally, we are given a dataset \(\mathcal{D} = \{(\vec{x}^{(i)}, y^{(i)}) \mid i = 1, \dots, n\}\), where \(\vec{x}^{(i)} \in \mathbb{N}^d\) encodes user \(i\)'s activity across \(d\) subreddits, and \(y^{(i)}\) is a binary label corresponding to one of three attributes:
\begin{itemize}
    \item \textbf{Year of Birth}: below vs. above the sample median,
    \item \textbf{Gender}: Male vs. Female,
    \item \textbf{Partisan Affiliation}: Democrat vs. Republican.
\end{itemize}
We evaluate models under two supervision settings: (1)\textit{true supervision}, which uses explicit self-disclosures from Reddit users, and (2)\textit{distant supervision}, which infers labels based on users’ participation in curated sets of “seed” subreddits associated with each class. The distant supervision procedure follows~\cite{waller2021quantifying} and is detailed in the Appendix. Briefly, a user is assigned a label for a given attribute if their participation in the corresponding set of subreddits exceeds that in the opposing set by a threshold margin, helping to reduce label noise.

Model performance is evaluated using average ROC AUC and F1 score across 100 bootstrap splits, with stratified sampling and class balancing performed using a random over-sampler.

\begin{figure*}[ht!]
\includegraphics[width=.99\textwidth]{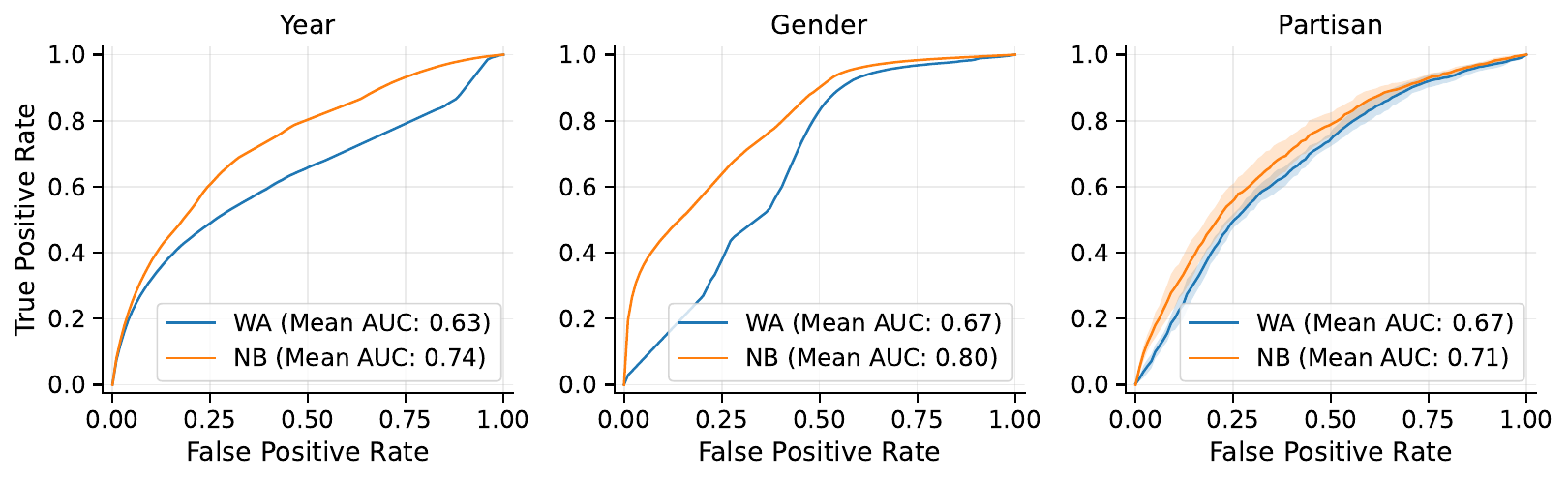}
\centering
\caption{ROC curves for each attribute (Year of Birth, Gender, Partisan Affiliation) and model (Naive Bayes (NB), \citeauthor{waller2021quantifying} (WA)).
Models are trained with \textbf{true supervision}, use random oversampling for class imbalance, and are evaluated via 10-fold stratified cross-validation.}
\label{fig:classif_true}
\end{figure*}

\spara{Simpler models outperform embedding-based models with supervised labeling.}
\Cref{fig:classif_true} shows that Naive Bayes-based models (NB) consistently outperform embedding-based models (WA) by a substantial margin in attributes when a sufficient amount of data is available.
This performance gap is particularly evident in the gender task, which benefits from the largest dataset size (424k declarations).
Naive Bayes achieves an average AUC of $0.80$ compared to $0.67$ for the WA model.

The datasets for gender and year are notably larger than those for the partisan attribute, with almost two orders of magnitude more declarations (\Cref{table:stats}), contributing to the performance of supervised methods in these cases.

\Cref{table:classif} shows details for models trained with \emph{true supervision}.
Naive Bayes consistently outperforms the Majority baseline by a large margin across all attributes and metrics.
Compared to the WA model, NB achieves significant improvements in all metrics, with margins ranging from $0.4$ to $1.8$, depending on the attribute.
Random Forest (RF), despite its higher computational cost compared to NB, demonstrates a competitive advantage only in the \textit{partisan} prediction task, where data scarcity dominates.
In this setting, RF slightly outperforms NB in ROC AUC, although the difference is within the error bars of the two classifiers.

\begin{figure*}[h!]
\centering
\includegraphics[width=.99\textwidth]{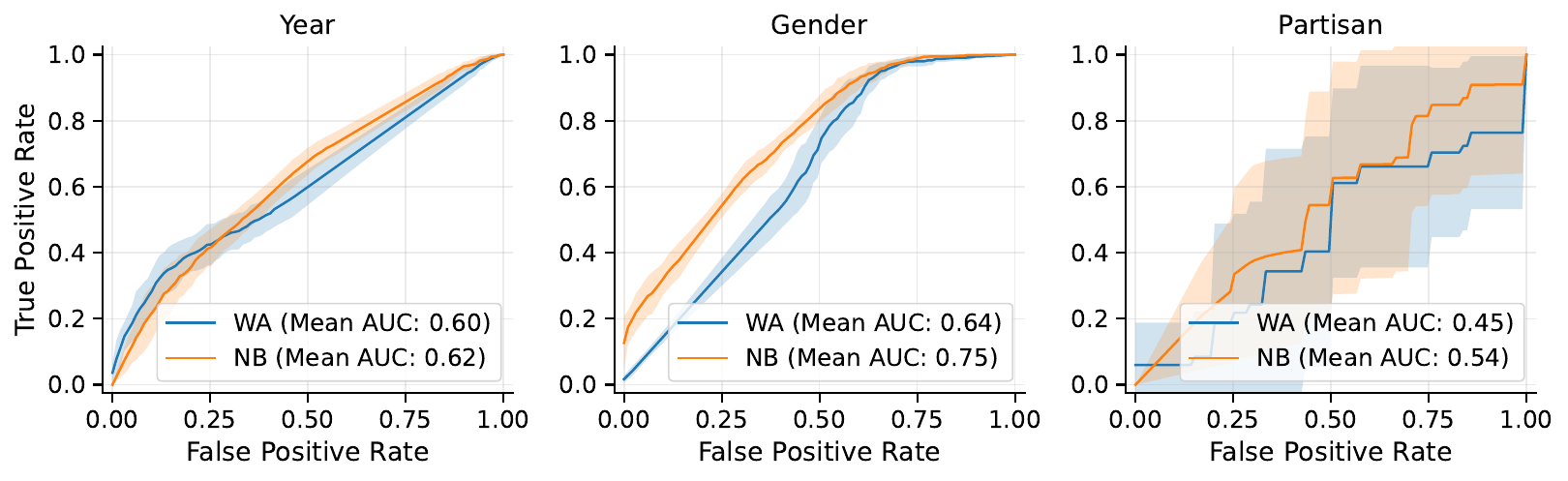}
\caption{ROC curves for each attribute (Year of Birth, Gender, Partisan Affiliation) and model (Naive Bayes (NB), \citeauthor{waller2021quantifying} (WA)).
Models are trained with \textbf{distant supervision}, use random oversampling for class imbalance, and are evaluated via 10-fold stratified cross-validation.}
\label{fig:classif_dist}
\end{figure*}

\spara{Simpler Models outperform embedding-based models with distant supervision.}
Figure~\ref{fig:classif_dist} demonstrates that Naive Bayes models (NB) outperform embedding-based methods (WA) when trained with labels derived solely from distant supervision.
These labels are generated based on participation in discriminative subreddits, as defined by \citet{waller2021quantifying}.
This result underscores the capability of Naive Bayes models to effectively capture sociodemographic patterns similar to those assumed in the embedding-based approach while retaining simplicity and interpretability.
Thus, the simpler approach is preferable even when self-declarations are not available.
However, Naive Bayes models trained under distant supervision show lower overall performance, higher variance, and narrower margins than their true-supervised counterparts and the WA model.
This reduced performance highlights the inherent limitations of distant supervision, which relies on noisier and less direct labels derived from coarse assumptions about user participation patterns.

\spara{True supervision labels are better than distant supervision.}
Training on declared labels yields better performance compared to distant supervision, as evidenced by the degraded scores observed from~\Cref{fig:classif_true} (true supervision) to~\Cref{fig:classif_dist} (distant supervision).
This performance gap arises from several factors.
First, declared labels are direct evidence for the target attributes and thus higher-quality training data.
Training and testing on the same type of data inherently offers an advantage due to consistency in data distributions.
Second, the dataset of declared labels is approximately two orders of magnitude larger than the dataset derived from distant supervision, which allows the supervised models to learn more robust patterns and generalize better.
Third, this approach leverages the full granularity of the participation and self-declaration data.

\section{Sociodemographic Quantification at Scale}
In demographic quantification, the goal is to estimate the \emph{prevalence} of a binary attribute within a user cohort, without classifying individuals. Formally, given a set of unlabeled feature vectors \(\{\vec{x}^{(j)}\}_{j=1}^m\), the task is to estimate the class distribution \(p(y)\) over the group for the three binary attributes \textit{Year of Birth}, \textit{Gender}, \textit{Partisan Affiliation}.
To support this task, we extend Naive Bayes with quantification wrappers (Classify and Count, Adjusted Classify and Count) and evaluate semi-supervised and log-normal variants.  
Performance is measured using Mean Absolute Error (MAE).
For each of 50 test sets---each sampled to approximate realistic class prevalences---we compute MAE across predictions, averaging results over 50 random seeds.

\spara{Simpler models outperform embedding-based in quantification with supervised labeling.}
\Cref{table:quant-true} presents the performance of the best-performing models and baseline models on the quantification task using true supervision.
The results demonstrate that simpler Bayesian models, such as Multinomial Naive Bayes (NB) and its variants, consistently outperform embedding-based approaches (WA).
For instance, the NB model achieves significantly lower mean absolute error (MAE) compared to the WA model and its variations across all attributes (gender, partisan, and year).

This advantage is particularly pronounced for the gender attribute, where the NB model not only reduces the MAE but also exhibits lower standard deviation, highlighting its robustness and effectiveness in quantification tasks.
Specifically, the MNB achieves absolute errors of approximately $11\%$, making it a reliable choice for this attribute.

While incorporating semi-supervised techniques or fitting a log-normal distribution to model user activity lead to incremental performance improvements, the computational overhead of these methods outweighs their benefits when considering the marginal gains relative to the error bars.

Additionally, adjustments to the WA model, such as applying adjusted classify-and-count (ACC) to refine conditional probabilities, result in better performance compared to the base WA classify-and-count (CC) method.
However, these improvements remain insufficient to match the performance of Naive Bayes models, with discrepancies as high as $15\%$ for gender.

\begin{figure*}[ht!]
\centering
\includegraphics[width=.9\textwidth]{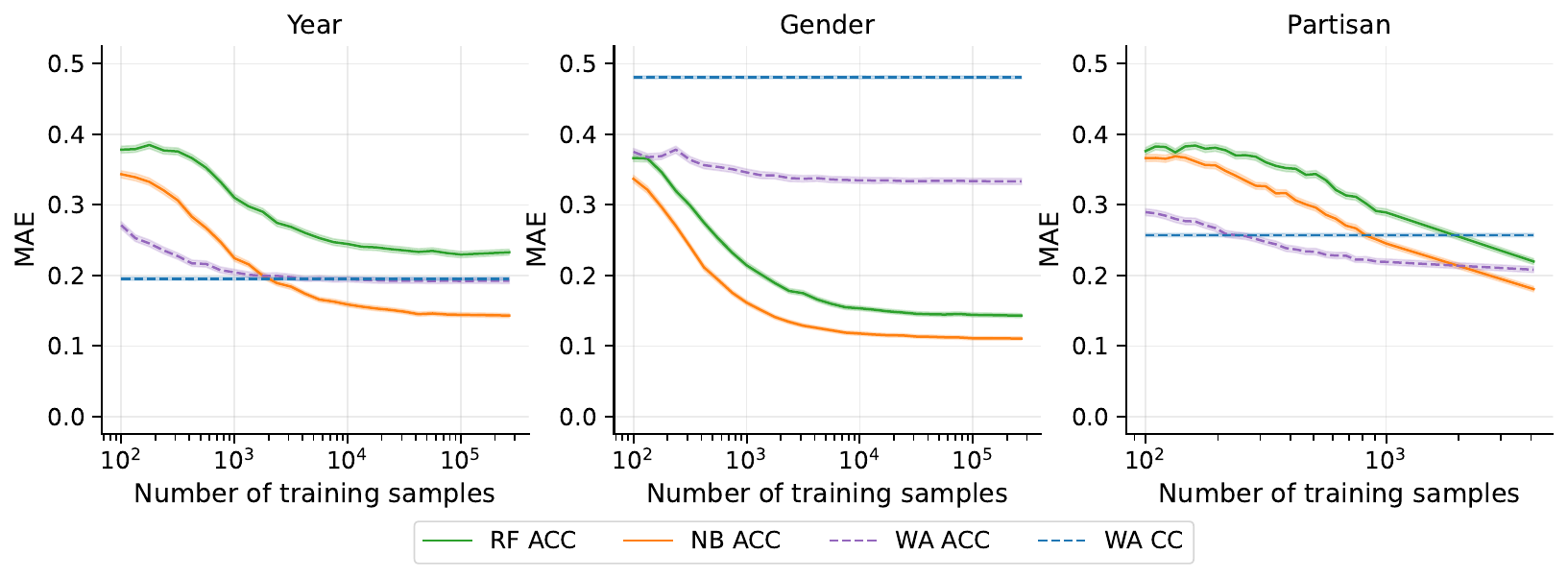}
\caption{Quantification curves. MAE obtained each method vs the number of training samples with true declared labels (true supervision).}
\label{fig:curves}
\end{figure*}

\begin{table}[t]
\caption{
Mean Absolute Error (MAE) for the quantification task with \textbf{true supervision} across three binary attributes (Year of Birth, Gender, Partisan Affiliation), reporting mean $\pm$ standard deviation over 50 randomized test samples. 
We evaluate multiple classifiers: 
\textbf{(i)} Majority baseline; 
\textbf{(ii)} WA model~\citep{waller2021quantifying}; 
\textbf{(iii)} Random Forest (RF) with 50 estimators and max depth 10; 
\textbf{(iv)} several Naive Bayes (NB) variants including log-normal activity modeling and semi-supervised EM training. 
All models are evaluated using the Adjusted Classify \& Count (ACC) wrapper, with WA and Majority also tested under the simpler Classify \& Count (CC). 
Best scores in bold.
}
\centering
\resizebox{\columnwidth}{!}{
\sisetup{table-format=1.3(1),table-text-alignment=center,round-mode=uncertainty,round-precision=3,uncertainty-mode=separate,retain-zero-uncertainty,detect-weight}
\newrobustcmd{\B}{\bfseries}
\begin{tabular}{l SSS}
\toprule
\textbf{Model}  & {\textbf{Year}} & {\textbf{Gender}}  & {\textbf{Partisan}} \\
\midrule
Majority ACC      & 0.532 \pm 0.088  & 0.319 \pm 0.172 & 0.344 \pm 0.175 \\
WA CC             & 0.195 \pm 0.079  & 0.481 \pm 0.070 & 0.257 \pm 0.082 \\
\midrule
WA ACC            & 0.192 \pm 0.137  & 0.333 \pm 0.147 & 0.208 \pm 0.146 \\
NB ACC            & \B 0.143 \pm 0.106  & 0.111 \pm 0.084 & \B 0.181 \pm 0.132 \\
NB logN ACC       & 0.144 \pm 0.107  & 0.111 \pm 0.085 & 0.181 \pm 0.130 \\
SS NB ACC         & \B 0.143 \pm 0.106  & \B 0.110 \pm 0.084 & 0.194 \pm 0.141 \\
SS NB logN ACC    & \B 0.143 \pm 0.106  & \B 0.110 \pm 0.084 & 0.190 \pm 0.138 \\
RF ACC            & 0.233 \pm 0.155  & 0.143 \pm 0.106 & 0.220 \pm 0.153 \\
\bottomrule
\end{tabular}

}
\label{table:quant-true}
\end{table}

\spara{Simpler bayesian models are  better than embedding-based with distant supervision.}
\Cref{table:quant-dist} shows that simpler Bayesian models, such as Multinomial Naive Bayes (MNB) and its semi-supervised variants, consistently outperform embedding-based models in quantification under distant supervision.
Notably, the best-performing Bayesian model achieves up to a $36\%$ improvement in gender quantification compared to the WA model with Classify and Count (CC).

The results also demonstrate the utility of assumptions inherent in distant supervision labeling.
For attributes such as year and gender, distant supervision produces comparable or better results while enabling the inclusion of a larger and more diverse user base in the training set.
This broader dataset captures more heterogeneous activity patterns, unlike self-declared labels, which may be biased toward users active in fewer subreddits or specific communities.

However, for the partisan attribute, distant supervision is more challenging.
Only more complex models, such as Random Forest, achieve acceptable performance, with MAE values below $0.2$.
These results highlight the limitations of simpler models for this attribute under distant supervision and emphasize the importance of leveraging self-declarations to validate the assumptions and ensure the reliability of distant labels.

Finally, these results underline the critical impact of direct sociodemographic proxies.
While distant supervision expands the training dataset, it inherently introduces noise and reduces performance due to its reliance on assumptions about user participation patterns.
In contrast, directly declared labels provide a more accurate and robust foundation for model training, which leads to superior performance.

\begin{table}[t]
\caption{
Mean Absolute Error (MAE) for the quantification task with \textbf{distant supervision} across three binary attributes (Year of Birth, Gender, Partisan Affiliation), reporting mean $\pm$ standard deviation over 50 randomized test samples. 
We evaluate multiple classifiers: 
\textbf{(i)} Majority baseline; 
\textbf{(ii)} WA model~\citep{waller2021quantifying}; 
\textbf{(iii)} Random Forest (RF) with 50 estimators and max depth 10; 
\textbf{(iv)} several Naive Bayes (NB) variants including log-normal activity modeling and semi-supervised EM training. 
All models are evaluated using the Adjusted Classify \& Count (ACC) wrapper, with WA and Majority also tested under the simpler Classify \& Count (CC). 
Best scores in bold.
}
\centering
\resizebox{\columnwidth}{!}{
\sisetup{table-format=1.3(1),table-text-alignment=center,round-mode=uncertainty,round-precision=3,uncertainty-mode=separate,retain-zero-uncertainty,detect-weight}
\newrobustcmd{\B}{\bfseries}

\begin{tabular}{lSSS}
\toprule
{\textbf{Model}}   & {\textbf{Year}} & {\textbf{Gender}} & {\textbf{Partisan}}\\
\midrule
Majority ACC       & 0.539 \pm 0.075 & 0.377 \pm 0.113 & 0.449 \pm 0.080	\\
WA CC              & 0.206 \pm 0.073 & 0.480 \pm 0.071 & 0.256 \pm 0.079		\\
\midrule
WA ACC             & 0.444 \pm 0.063 & 0.472 \pm 0.114 & 0.434 \pm 0.116		\\
NB ACC             & 0.094 \pm 0.079 & 0.112 \pm 0.084 & 0.407 \pm 0.084		\\
NB logN ACC        &\B 0.084 \pm 0.068 &\B 0.110 \pm 0.083 & 0.410 \pm 0.084\\
SS NB ACC          & 0.086 \pm 0.065 &\B 0.110 \pm 0.083 & 0.408 \pm 0.085	\\
SS NB logN ACC     & 0.086 \pm 0.070 & 0.113 \pm 0.087 & 0.408 \pm 0.084	\\
RF ACC             & 0.536 \pm 0.066 & 0.495 \pm 0.074 &\B 0.156 \pm 0.106	\\
\bottomrule
\end{tabular}}
\label{table:quant-dist}
\end{table}

\spara{Supervised models require at least 1k data points to outperform pretrained models.}
\Cref{fig:curves} illustrates the reduction in MAE as a function of the number of training samples for the various models.
The WA model (CC) relies on classification followed by counting the proportions of each class for the given binary attribute.
As it does not depend on the size of the training dataset, its performance remains constant across all data points.
In contrast, all other methods are trained on datasets with true supervision, and the quantifier wrappers (ACC) further refine these models by fitting conditional distributions in the training set to adjust the class proportions.

For \emph{year} and \emph{partisan}, the WA Model converges quickly but reaches a performance plateau that falls short of the Naive Bayes models.
Similarly, Random Forest exhibits consistently higher MAE than Naive Bayes across all training set sizes. 
On \emph{gender} in particular, results underscore the importance of quantifiers.
Training quantifiers to estimate conditional probabilities is fundamental for accurate quantification of this attribute, as shown by the poor performance of the Classify and Count (CC) technique.

Finally, the size of the training data emerges as a critical factor in model performance.
Across all attributes, a minimum of \num{1000} training samples is required for supervised models to surpass the performance of the pretrained WA model.

\section{Model Transparency Enables Auditing and Interpretability}
\begin{figure*}[ht!]
\centering
\includegraphics[width=.99\textwidth]{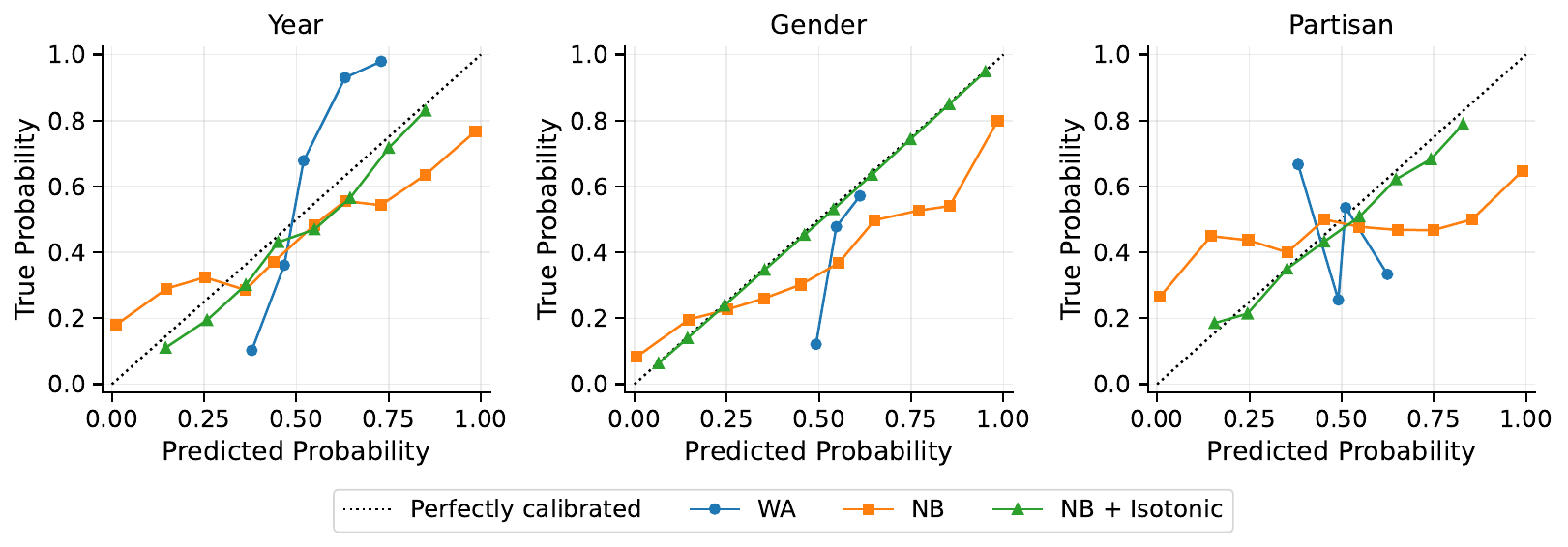}
\caption{Calibration curves for the different attributes, showing the alignment of prediction scores with true probabilities.}
\label{fig:calibration-year}
\end{figure*}

By design, Bayesian models inherently provide interpretable characteristics, making them ideal for tasks requiring transparency and explainability.
Here, we discuss key interpretability aspects derived from our models.

\spara{Feature Importance.}
The interpretability of a Naive Bayes (NB) model stems from its ability to assign log-odds weights to features, reflecting their average importance in predictions.
To illustrate how the model enables researchers to derive interpretable insights from feature importance and foster a deeper understanding of sociodemographic patterns, we analyze the log odds and their standard deviations to identify the most influential features for different attributes.

For \emph{gender}, the key features include \texttt{SkincareAddiction}, \texttt{raisedbynarcissists}, and \texttt{buildapc}, as shown in~\Cref{fig:importance_combined} (b).
For \emph{partisan affiliation}, significant subreddits include \texttt{Conservative}, \texttt{The\_Donald}, and \texttt{SandersForPresident}, as depicted in~\Cref{fig:importance_combined} (c).
For \emph{year of birth}, features such as \texttt{teenagers}, \texttt{Minecraft}, and \texttt{dankmemes} dominate in determining the year attribute, as illustrated in~\Cref{fig:importance_combined} (a).

This analysis showcases the practical utility of Naive Bayes models in uncovering interpretable feature importance, enabling researchers to better understand the sociodemographic behaviors embedded in the data.


\spara{Calibration.}
An important aspect of a predictive model is the ability to interpret its predicted scores for each data point as probabilities, which enables meaningful and actionable insights in the downstream tasks.
To maximize coverage of Reddit's user base, we use the most popular subreddits as the feature space.
However, it introduces challenges such as feature redundancy and overconfidence in Naive Bayes predictions, potentially compromising the reliability of the model's predictions.

To address these issues, we use isotonic regression to calibrate the prediction scores (\Cref{fig:calibration-year}).
Calibration ensures that the predicted scores align with true probabilities, and enhances both the interpretability and reliability of the model.
Notably, Naive Bayes provides a well-calibrated baseline compared to more complex methods, and its inherent simplicity makes it easier to calibrate effectively.
This advantage allows Naive Bayes to maintain accurate probability estimates even in scenarios with large and redundant feature spaces.

\spara{Prediction Intervals.}
In the quantification task, data points are independently but not identically distributed, as each data point is sampled according to a Bernoulli distribution with a success probability equal to its calibrated probability score.
Calibration ensures that these scores reflect true probabilities, making them reliable inputs for further statistical modeling.

Consequently, the Poisson-Binomial distribution, which generalizes the Binomial distribution for non-identical probabilities, describes the joint distribution of the data.
By leveraging the calibrated scores, we can use the known second moment of the Poisson-Binomial distribution to estimate prediction intervals for the quantification measure.
These intervals provide a reliable measure of uncertainty around the class proportions, offering easy-to-interpret insights into the variability of the predicted outcomes.

\spara{Robustness.}
To evaluate the robustness of the classifiers under varying conditions, we tested their performance by modifying user selection thresholds.
Specifically, we filtered users based on extreme score values, selecting those with scores either greater than a threshold $\tau$ or less than 1 - $\tau$, where $\tau \in [0, 1]$.
This approach examines how classifier performance responds to restricting the test set to users with high confidence predictions.

All classifiers exhibited strong stability across most threshold values.
Performance remained consistent, with minimal variation until $\tau$ dropped below $0.1$.
At this point, scores varied by $6\%$ to $12\%$, depending on the attribute and classifier.
These findings suggest that classifiers are robust under varying confidence thresholds.

\section{Discussion: Toward Ethical and Auditable Social Data Science}
\label{sec:discussion}
Sociodemographic characterization enables researchers to understand how online communities reflect social identities and to model the complex interactions between individuals and their environments. It is also an invaluable tool for investigating key societal questions.
However, research on Reddit has lacked systematic evaluations and clear guidelines to identify effective methodologies.
This study addresses this gap by introducing a reliable, transparent, and efficient framework for sociodemographic inference.
We offer actionable best-practice insights for CSS research on Reddit and remark on the coverage, interpretability, reliability, and scalability of models.

Our pipeline spans data collection, modeling, and task execution.
Thanks to the dataset we collect of over \num{850000} self-declarations on age, gender, and partisan affiliation, we can train and evaluate supervised learning methods with high-quality data.
This approach avoids reliance on assumed inferred characteristics and preserves the authenticity of the ground-truth data.
Moreover, the regex-based data collection process is scalable and robust, thus mitigating labeling costs and reducing potential annotation inconsistencies.

Our approach also highlights the value of the Bayesian framework for sociodemographic inference.
This framework not only yields strong empirical performance, but also enables transparency through calibrated outputs and subreddit-level feature importance
It supports further analysis using log odds and enables uncertainty quantification through prediction intervals, thus offering researchers a reliable measure of the robustness of the model.
These features should empower researchers to conduct their analyses with greater confidence.

Our findings caution researchers against overly engineered and unnecessarily complex models.
Instead, the results emphasize the effectiveness of simpler approaches that respect users’ self-declared information while maintaining methodological rigor.
This balance between simplicity, interpretability, and ethical transparency is critical for fostering trust in computational social science research.

Finally, our evaluation of classification and quantification tasks provides researchers with clear use cases for both user-level and population-level analysis.
The quantification task, in particular, emphasizes aggregated insights over individual predictions, thus offering a privacy-conscious alternative that reduces risks associated with exposing sensitive user information.
The models analyzed in this work achieve both efficiency and accuracy on both tasks, which makes them a robust choice for sociodemographic inference.

Despite the usefulness of our work, it is important to acknowledge some of its limitations.
(\emph{i}) \textit{Risks of Mislabeling.}
Mislabeling in population studies poses significant risks, including flawed analyses, spurious conclusions, and potentially harmful policy recommendations. Our findings emphasize the need for robust, interpretable pipelines to mitigate these risks and ensure the reliability of sociodemographic insights.
(\emph{ii}) \textit{Binary attributes.}
To simplify modeling, we reduced all attributes to binary pairs.
On gender and political affiliation, such a simplification might reduce the complexity of the world and reduce the visibility of minorities.
While the data did not present enough examples to consider other cases, it is important to be conscious of this limitation.
(\emph{iii}) \textit{Dynamic attributes.}
For similar reasons, we excluded ambiguous and inconsistent user declarations. However, this approach opens opportunities for future research into dynamic disclosures, particularly in analyzing how partisan declarations evolve in response to exogenous events.
(\emph{iv}) \textit{Integration of multiple proxies.}
Our study highlights the strengths and limitations of different proxies, including self-declared labels and distant supervision.
Nevertheless, integrating multiple proxies into a unified model remains underexplored and represents a promising avenue for future research.
(\emph{v}) \textit{Data sources.}
Finally, in this work, we focused on the possibilities offered by subreddit participation data on demographic inference, excluding from our scope alternative sources, e.g., textual comments made by users.
While such alternatives are computationally more expensive and less interpretable, it would be important to quantify their effectiveness in future work.



\section{Data availability}
The data underlying this article are available at:
\url{https://github.com/FedericoCinus/reddit-fabric}, with the dataset itself hosted at:
\url{https://drive.google.com/file/d/1vy_g9B6WM7BfO9uDUzrC_O8rAQfQCuIu/view}.

\bibliographystyle{plainnat}
\bibliography{reference}

\begin{thebibliography}{45}
\providecommand{\natexlab}[1]{#1}
\providecommand{\url}[1]{\texttt{#1}}
\expandafter\ifx\csname urlstyle\endcsname\relax
  \providecommand{\doi}[1]{doi: #1}\else
  \providecommand{\doi}{doi: \begingroup \urlstyle{rm}\Url}\fi

\bibitem[Bach et~al.(2021)Bach, Kern, Amaya, Keusch, Kreuter, Hecht, and
  Heinemann]{bach2021predicting}
Ruben~L Bach, Christoph Kern, Ashley Amaya, Florian Keusch, Frauke Kreuter, Jan
  Hecht, and Jonathan Heinemann.
\newblock Predicting voting behavior using digital trace data.
\newblock \emph{Social Science Computer Review}, 39\penalty0 (5):\penalty0
  862--883, 2021.

\bibitem[Balsamo et~al.(2019)Balsamo, Bajardi, and
  Panisson]{balsamo2019firsthand}
Duilio Balsamo, Paolo Bajardi, and Andr{\'e} Panisson.
\newblock Firsthand opiates abuse on social media: monitoring geospatial
  patterns of interest through a digital cohort.
\newblock In \emph{The World Wide Web Conference}, pages 2572--2579, 2019.

\bibitem[Balsamo et~al.(2023)Balsamo, Bajardi, De~Francisci~Morales, Monti, and
  Schifanella]{balsamo2023pursuit}
Duilio Balsamo, Paolo Bajardi, Gianmarco De~Francisci~Morales, Corrado Monti,
  and Rossano Schifanella.
\newblock The {Pursuit} of {Peer} {Support} for {Opioid} {Use} {Recovery} on
  {Reddit}.
\newblock \emph{International AAAI Conference on Web and Social Media},
  17:\penalty0 12--23, June 2023.
\newblock ISSN 2334-0770, 2162-3449.
\newblock \doi{10.1609/icwsm.v17i1.22122}.
\newblock URL \url{https://ojs.aaai.org/index.php/ICWSM/article/view/22122}.

\bibitem[Barbosa and Chen(2019)]{barbosa2019rehumanized}
Nat{\~a}~M Barbosa and Monchu Chen.
\newblock Rehumanized crowdsourcing: A labeling framework addressing bias and
  ethics in machine learning.
\newblock In \emph{2019 CHI Conference on Human Factors in Computing Systems},
  pages 1--12, 2019.

\bibitem[Baumgartner et~al.(2020)Baumgartner, Zannettou, Keegan, Squire, and
  Blackburn]{baumgartner2020pushshift}
Jason Baumgartner, Savvas Zannettou, Brian Keegan, Megan Squire, and Jeremy
  Blackburn.
\newblock The pushshift reddit dataset.
\newblock In \emph{international AAAI conference on web and social media},
  volume~14, pages 830--839, 2020.

\bibitem[Betti et~al.(2025)Betti, Bajardi, and
  De~Francisci~Morales]{betti2025moral}
Lorenzo Betti, Paolo Bajardi, and Gianmarco De~Francisci~Morales.
\newblock Moral judgments in online discourse are not biased by gender.
\newblock \emph{Scientific Reports}, 15\penalty0 (1):\penalty0 21555, 2025.

\bibitem[Capozzi et~al.(2021)Capozzi, De~Francisci~Morales, Mejova, Monti,
  Panisson, and Paolotti]{capozzi2021clandestino}
Arthur Capozzi, Gianmarco De~Francisci~Morales, Yelena Mejova, Corrado Monti,
  Andr{\'e} Panisson, and Daniela Paolotti.
\newblock Clandestino or rifugiato? anti-immigration facebook ad targeting in
  italy.
\newblock In \emph{2021 CHI Conference on Human Factors in Computing Systems},
  pages 1--15, 2021.

\bibitem[Chawla et~al.(2002)Chawla, Bowyer, Hall, and
  Kegelmeyer]{chawla2002smote}
Nitesh~V Chawla, Kevin~W Bowyer, Lawrence~O Hall, and W~Philip Kegelmeyer.
\newblock Smote: synthetic minority over-sampling technique.
\newblock \emph{Journal of artificial intelligence research}, 16:\penalty0
  321--357, 2002.

\bibitem[Chew et~al.(2021)Chew, Kery, Baum, Bukowski, Kim, Navarro,
  et~al.]{chew2021predicting}
Robert Chew, Caroline Kery, Laura Baum, Thomas Bukowski, Annice Kim, Mario
  Navarro, et~al.
\newblock Predicting age groups of reddit users based on posting behavior and
  metadata: classification model development and validation.
\newblock \emph{JMIR Public Health and Surveillance}, 7\penalty0 (3):\penalty0
  e25807, 2021.

\bibitem[Cinelli et~al.(2021)Cinelli, De~Francisci~Morales, Galeazzi,
  Quattrociocchi, and Starnini]{cinelli2021echo}
Matteo Cinelli, Gianmarco De~Francisci~Morales, Alessandro Galeazzi, Walter
  Quattrociocchi, and Michele Starnini.
\newblock The {Echo} {Chamber} {Effect} on {Social} {Media}.
\newblock \emph{Proceedings of the National Academy of Sciences}, 118\penalty0
  (9):\penalty0 e2023301118, 2021.

\bibitem[Colacrai et~al.(2024)Colacrai, Cinus, De~Francisci~Morales, and
  Starnini]{colacrai2024navigating}
Ernesto Colacrai, Federico Cinus, Gianmarco De~Francisci~Morales, and Michele
  Starnini.
\newblock Navigating multidimensional ideologies with reddit's political
  compass: Economic conflict and social affinity.
\newblock In \emph{ACM on Web Conference 2024}, pages 2582--2593, 2024.

\bibitem[De~Candia et~al.(2022)De~Candia, De~Francisci~Morales, Monti, and
  Bonchi]{de2022social}
Sara De~Candia, Gianmarco De~Francisci~Morales, Corrado Monti, and Francesco
  Bonchi.
\newblock Social norms on reddit: A demographic analysis.
\newblock In \emph{14th ACM Web Science Conference 2022}, pages 139--147, 2022.

\bibitem[del Rio-Chanona et~al.(2023)del Rio-Chanona, Hermida-Carrillo,
  Sepahpour-Fard, Sun, Topinkova, and Nedelkoska]{del2023mental}
R~Maria del Rio-Chanona, Alejandro Hermida-Carrillo, Melody Sepahpour-Fard,
  Luning Sun, Renata Topinkova, and Ljubica Nedelkoska.
\newblock Mental health concerns precede quits: shifts in the work discourse
  during the covid-19 pandemic and great resignation.
\newblock \emph{EPJ Data Science}, 12\penalty0 (1):\penalty0 49, 2023.

\bibitem[Eickhoff(2018)]{eickhoff2018cognitive}
Carsten Eickhoff.
\newblock Cognitive biases in crowdsourcing.
\newblock In \emph{eleventh ACM international conference on web search and data
  mining}, pages 162--170, 2018.

\bibitem[Fan et~al.(2023)Fan, Su, Sun, Noyman, Zhang, Pentland, and
  Moro]{fan2023diversity}
Zhuangyuan Fan, Tianyu Su, Maoran Sun, Ariel Noyman, Fan Zhang, Alex Pentland,
  and Esteban Moro.
\newblock Diversity beyond density: Experienced social mixing of urban streets.
\newblock \emph{PNAS nexus}, 2\penalty0 (4):\penalty0 pgad077, 2023.

\bibitem[Giorgi et~al.(2022)Giorgi, Lynn, Gupta, Ahmed, Matz, Ungar, and
  Schwartz]{giorgi2022correcting}
Salvatore Giorgi, Veronica~E Lynn, Keshav Gupta, Farhan Ahmed, Sandra Matz,
  Lyle~H Ungar, and H~Andrew Schwartz.
\newblock Correcting sociodemographic selection biases for population
  prediction from social media.
\newblock In \emph{International AAAI Conference on Web and Social Media},
  volume~16, pages 228--240, 2022.

\bibitem[Gjurkovi{\'c} et~al.(2021)Gjurkovi{\'c}, Karan, Vukojevi{\'c},
  Bo{\v{s}}njak, and {\v{S}}najder]{gjurkovic2021pandora}
Matej Gjurkovi{\'c}, Mladen Karan, Iva Vukojevi{\'c}, Mihaela Bo{\v{s}}njak,
  and Jan {\v{S}}najder.
\newblock Pandora talks: Personality and demographics on reddit.
\newblock In \emph{Ninth International Workshop on Natural Language Processing
  for Social Media}, pages 138--152, 2021.

\bibitem[Gonz{\'a}lez et~al.(2017)Gonz{\'a}lez, Casta{\~n}o, Chawla, and
  Coz]{gonzalez2017review}
Pablo Gonz{\'a}lez, Alberto Casta{\~n}o, Nitesh~V Chawla, and Juan Jos{\'e}~Del
  Coz.
\newblock A review on quantification learning.
\newblock \emph{ACM Computing Surveys (CSUR)}, 50\penalty0 (5):\penalty0 1--40,
  2017.

\bibitem[Hamilton(2011)]{hamilton2011all}
James~T Hamilton.
\newblock All the news that's fit to sell.
\newblock In \emph{All the News That's Fit to Sell}. Princeton University
  Press, 2011.

\bibitem[He et~al.(2008)He, Bai, Garcia, and Li]{he2008adasyn}
Haibo He, Yang Bai, Edwardo~A Garcia, and Shutao Li.
\newblock Adasyn: Adaptive synthetic sampling approach for imbalanced learning.
\newblock In \emph{2008 IEEE international joint conference on neural networks
  (IEEE world congress on computational intelligence)}, pages 1322--1328. Ieee,
  2008.

\bibitem[Hettiachchi et~al.(2021)Hettiachchi, Sanderson, Goncalves, Hosio,
  Kazai, Lease, Schaekermann, and Yilmaz]{hettiachchi2021investigating}
Danula Hettiachchi, Mark Sanderson, Jorge Goncalves, Simo Hosio, Gabriella
  Kazai, Matthew Lease, Mike Schaekermann, and Emine Yilmaz.
\newblock Investigating and mitigating biases in crowdsourced data.
\newblock In \emph{Companion Publication of the 2021 Conference on Computer
  Supported Cooperative Work and Social Computing}, pages 331--334, 2021.

\bibitem[Kim et~al.(2025)Kim, Veselovsky, and Anderson]{kim2025capturing}
Rachel~Minyoung Kim, Veniamin Veselovsky, and Ashton Anderson.
\newblock Capturing dynamics in online public discourse: A case study of
  universal basic income discussions on reddit.
\newblock In \emph{Proceedings of the International AAAI Conference on Web and
  Social Media}, volume~19, pages 1021--1037, 2025.

\bibitem[Kittur et~al.(2008)Kittur, Chi, and Suh]{kittur2008crowdsourcing}
Aniket Kittur, Ed~H Chi, and Bongwon Suh.
\newblock Crowdsourcing user studies with mechanical turk.
\newblock In \emph{SIGCHI conference on human factors in computing systems},
  pages 453--456, 2008.

\bibitem[Klein et~al.(2019)Klein, Clutton, and Dunn]{klein2019pathways}
Colin Klein, Peter Clutton, and Adam~G Dunn.
\newblock Pathways to conspiracy: The social and linguistic precursors of
  involvement in reddit's conspiracy theory forum.
\newblock \emph{PloS one}, 14\penalty0 (11):\penalty0 e0225098, 2019.

\bibitem[Lee and Chyi(2014)]{lee2014newsworthy}
Angela~M Lee and Hsiang~Iris Chyi.
\newblock When newsworthy is not noteworthy: Examining the value of news from
  the audience's perspective.
\newblock \emph{Journalism studies}, 15\penalty0 (6):\penalty0 807--820, 2014.

\bibitem[Lokala et~al.(2022)Lokala, Srivastava, Dastidar, Chakraborty, Akhtar,
  Panahiazar, and Sheth]{lokala2022computational}
Usha Lokala, Aseem Srivastava, Triyasha~Ghosh Dastidar, Tanmoy Chakraborty,
  Md~Shad Akhtar, Maryam Panahiazar, and Amit Sheth.
\newblock A computational approach to understand mental health from reddit:
  knowledge-aware multitask learning framework.
\newblock In \emph{International AAAI Conference on Web and Social Media},
  volume~16, pages 640--650, 2022.

\bibitem[Medvedev et~al.(2019)Medvedev, Lambiotte, and
  Delvenne]{medvedev2019anatomy}
Alexey~N Medvedev, Renaud Lambiotte, and Jean-Charles Delvenne.
\newblock The anatomy of reddit: An overview of academic research.
\newblock \emph{Dynamics on and of Complex Networks III: Machine Learning and
  Statistical Physics Approaches 10}, pages 183--204, 2019.

\bibitem[Mok et~al.(2023)Mok, Inzlicht, and Anderson]{mok2023echo}
Lillio Mok, Michael Inzlicht, and Ashton Anderson.
\newblock Echo tunnels: Polarized news sharing online runs narrow but deep.
\newblock In \emph{International AAAI Conference on Web and Social Media},
  volume~17, pages 662--673, 2023.

\bibitem[Monti et~al.(2022)Monti, Aiello, De~Francisci~Morales, and
  Bonchi]{monti2022language}
Corrado Monti, Luca~Maria Aiello, Gianmarco De~Francisci~Morales, and Francesco
  Bonchi.
\newblock The language of opinion change on social media under the lens of
  communicative action.
\newblock \emph{Scientific Reports}, 12\penalty0 (1):\penalty0 17920, 2022.

\bibitem[Monti et~al.(2023)Monti, D'Ignazi, Starnini, and
  De~Francisci~Morales]{monti2023evidence}
Corrado Monti, Jacopo D'Ignazi, Michele Starnini, and Gianmarco
  De~Francisci~Morales.
\newblock Evidence of demographic rather than ideological segregation in news
  discussion on reddit.
\newblock In \emph{ACM Web Conference 2023}, pages 2777--2786, 2023.

\bibitem[Moreo et~al.(2021)Moreo, Esuli, and Sebastiani]{moreo2021quapy}
Alejandro Moreo, Andrea Esuli, and Fabrizio Sebastiani.
\newblock Quapy: a python-based framework for quantification.
\newblock In \emph{30th ACM International Conference on Information \&
  Knowledge Management}, pages 4534--4543, 2021.

\bibitem[{Nature Editorial}(2021)]{nature2021digital}
{Nature Editorial}.
\newblock The powers and perils of using digital data to understand human
  behaviour.
\newblock \emph{Nature}, 595:\penalty0 149--150, 2021.
\newblock \doi{10.1038/d41586-021-01736-y}.
\newblock URL \url{https://www.nature.com/articles/d41586-021-01736-y}.
\newblock Editorial.

\bibitem[Nigam et~al.(2000)Nigam, McCallum, Thrun, and Mitchell]{nigam2000text}
Kamal Nigam, Andrew~Kachites McCallum, Sebastian Thrun, and Tom Mitchell.
\newblock Text classification from labeled and unlabeled documents using em.
\newblock \emph{Machine learning}, 39:\penalty0 103--134, 2000.

\bibitem[Rivas et~al.(2020)Rivas, Sadah, Guo, Hristidis,
  et~al.]{rivas2020classification}
Ryan Rivas, Shouq~A Sadah, Yuhang Guo, Vagelis Hristidis, et~al.
\newblock Classification of health-related social media posts: Evaluation of
  post content--classifier models and analysis of user demographics.
\newblock \emph{JMIR Public Health and Surveillance}, 6\penalty0 (2):\penalty0
  e14952, 2020.

\bibitem[Rollo et~al.(2022)Rollo, De~Francisci~Morales, Monti, and
  Panisson]{rollo2022communities}
Cesare Rollo, Gianmarco De~Francisci~Morales, Corrado Monti, and Andr{\'e}
  Panisson.
\newblock Communities, {Gateways}, and {Bridges}: {Measuring} {Attention}
  {Flow} in the {Reddit} {Political} {Sphere}.
\newblock In Frank Hopfgartner, Kokil Jaidka, Philipp Mayr, Joemon Jose, and
  Jan Breitsohl, editors, \emph{Social {Informatics}}, volume 13618 of
  \emph{{SocInfo}}, pages 3--19. Springer, [best paper award], 2022.
\newblock ISBN 978-3-031-19096-4 978-3-031-19097-1.

\bibitem[Sasse et~al.(2024)Sasse, Mahabir, Gkountouna, Crooks, and
  Croitoru]{sasse2024understanding}
Kuleen Sasse, Ron Mahabir, Olga Gkountouna, Andrew Crooks, and Arie Croitoru.
\newblock Understanding the determinants of vaccine hesitancy in the united
  states: A comparison of social surveys and social media.
\newblock \emph{Plos one}, 19\penalty0 (6):\penalty0 e0301488, 2024.

\bibitem[Shahrokh~Esfahani and Dougherty(2014)]{shahrokh2014effect}
Mohammad Shahrokh~Esfahani and Edward~R Dougherty.
\newblock Effect of separate sampling on classification accuracy.
\newblock \emph{Bioinformatics}, 30\penalty0 (2):\penalty0 242--250, 2014.

\bibitem[Shorten and Khoshgoftaar(2019)]{shorten2019survey}
Connor Shorten and Taghi~M Khoshgoftaar.
\newblock A survey on image data augmentation for deep learning.
\newblock \emph{Journal of big data}, 6\penalty0 (1):\penalty0 1--48, 2019.

\bibitem[Sristy and Somayajulu(2012)]{sristy2012semi}
Nagesh~Bhattu Sristy and DVLN Somayajulu.
\newblock Semi-supervised learning of naive bayes classifier with feature
  constraints.
\newblock In \emph{First International Workshop on Optimization Techniques for
  Human Language Technology}, pages 65--78, 2012.

\bibitem[Stritch et~al.(2017)Stritch, Pedersen, and
  Taggart]{stritch2017opportunities}
Justin~M Stritch, Mogens~Jin Pedersen, and Gabel Taggart.
\newblock The opportunities and limitations of using mechanical turk (mturk) in
  public administration and management scholarship.
\newblock \emph{International Public Management Journal}, 20\penalty0
  (3):\penalty0 489--511, 2017.

\bibitem[Tadesse et~al.(2019)Tadesse, Lin, Xu, and Yang]{tadesse2019detection}
Michael~M Tadesse, Hongfei Lin, Bo~Xu, and Liang Yang.
\newblock Detection of depression-related posts in reddit social media forum.
\newblock \emph{Ieee Access}, 7:\penalty0 44883--44893, 2019.

\bibitem[Vasilev(2018)]{vasilev2018inferring}
Evgenii Vasilev.
\newblock \emph{Inferring gender of reddit users}.
\newblock PhD thesis, Universit{\"a}t Koblenz, 2018.

\bibitem[Waller and Anderson(2021)]{waller2021quantifying}
Isaac Waller and Ashton Anderson.
\newblock Quantifying social organization and political polarization in online
  platforms.
\newblock \emph{Nature}, 600\penalty0 (7888):\penalty0 264--268, 2021.

\bibitem[Wang and Jurgens(2018)]{wang2018s}
Zijian Wang and David Jurgens.
\newblock It's going to be okay: Measuring access to support in online
  communities.
\newblock In \emph{Proceedings of the 2018 Conference on Empirical Methods in
  Natural Language Processing}, pages 33--45, 2018.

\bibitem[Zhang et~al.(2016)Zhang, Wu, and Sheng]{zhang2016learning}
Jing Zhang, Xindong Wu, and Victor~S Sheng.
\newblock Learning from crowdsourced labeled data: a survey.
\newblock \emph{Artificial Intelligence Review}, 46:\penalty0 543--576, 2016.

\end{thebibliography}

\clearpage
\section*{Materials and Methods}
\label{sec:mat}
\setcounter{figure}{0}
\renewcommand{\thefigure}{A\arabic{figure}}

\subsection{Models Overview}
We frame sociodemographic classification as the task of predicting binary attributes (e.g., gender, age cohort, partisanship) from users’ subreddit participation patterns. 

We evaluate a range of models that reflect common practices in the field, including Bayesian models, embedding-based models~\citep{waller2021quantifying}, decision tree-based models, semi-supervised approaches, and activity-modeling techniques.

\begin{itemize}
    \item \textbf{WA Model \citep{waller2021quantifying}:} Each subreddit has a pre-trained embedding represented in a unified, high-dimensional vector space.
    For each attribute (age, gender, affluence, partisanship), a one-dimensional axis is defined by taking ``extreme'' subreddits as poles for the axis (called ``seeds'' in the original work).
    Other subreddits can then be projected onto the one-dimensional axis for each attribute.
    We assign a z-score to each subreddit for each dimension.
    Users can also be projected if represented as a weighted combination of the subreddits where they participated.
    A user's z-score for an attribute is the weighted average of the subreddit z-scores, weighted by the number of comments the user posted in each subreddit.
    The resulting z-score is then used to predict the user's class for the attribute.

    \item \textbf{Random Forest:} Random Forest is an ensemble learning method for classification based on decision trees. It is widely used for tabular data due to its flexibility, strong performance, and resistance to multicollinearity.

    \item \textbf{Bayesian Modeling:} Multinomial Naive Bayes (NB) is a standard Bayesian model that assumes feature independence and provides transparent, interpretable predictions.
    It is fast to train, with closed-form estimators for its parameters.
    We extend the semi-supervised Multinomial Naive Bayes model introduced by \citet{nigam2000text} and adapt it to our task with some promising extensions to Naive Bayes.

    \begin{enumerate}
        \item \emph{Semi-Supervised Naive Bayes} (\texttt{SS NB}): incorporates the Expectation-Maximization (EM) algorithm to enable semi-supervised learning, allowing the model to leverage both labeled and unlabeled data~\cite{sristy2012semi}.
        \item \emph{Log-Normal Naive Bayes} (\texttt{logN NB}): models a class-dependent user activity by assuming a log-normal distribution. Activity is measured as the $L_1$-norm of the feature vector. This approach allows for variations in user engagement.
        \item \emph{Semi-Supervised Log-Normal Naive Bayes} (\texttt{SS logN NB}): combines the semi-supervised EM approach with the log-normal activity model, thereby enhancing performance in scenarios with limited labeled data.
    \end{enumerate}
    More details are in the following subsection.
\end{itemize}


\subsection{Bayesian Modeling}
\label{sec:app_bayes}
We extend the semi-supervised Multinomial Naive Bayes (MNB) model introduced by~\citet{nigam2000text} by incorporating the conditional probability of subreddit activations given the class label.
First, we review the extended model, followed by a description of the Expectation-Maximization (EM) algorithm used for semi-supervised learning in Naive Bayes as outlined by~\citet{nigam2000text}.

\vspace{2mm}

Each data sample in the dataset $\mathcal{D}=\{(\vec{x}^{(i)}, y^{(i)}) \mid i=1,\ldots,n\}$ consists of an observed sequence of subreddit activations  $\Vec{x}^{(i)} \in \mathbb{N}^d$, where $d$ is the number of subreddits, and a class label $y^{(i)}\in \{0, \ldots,  k-1\}$ for user $i$.

In our case study, $k = 2$ for each attribute (\textit{Year of Birth}, \textit{Gender}, and \textit{Partisan Affiliation}), but we present the model in its general form.
Missing labels are denoted as $y^{(i)}=-1$.
The observed user’s total activity is given by $a^{(i)}=\lvert\vec{x}^{(i)}\rvert_1$.

\vspace{2mm}

In a naive sense, each user's activation over the set of $d$ subreddits is assumed to be independent.
In addition, the probabilities of activity $a$ and activations $\Vec{x}$ we assume to be conditionally independent given the class $y$.
Hence, we define the joint likelihood of the unsupervised as follows, by setting marginalizing out the labels
\begin{align*}
     \mathcal{L}(\vec{\Theta}) &= \prod_{i=1}^{n} p(\vec{x}^{(i)})\\
     &=  \prod_{i=1}^{n}  \sum_{y=0}^{k-1} p(\vec{x}^{(i)}, y) \\
     &=  \prod_{i=1}^{n}  \sum_{y=0}^{k-1}\left( p(y) p(a^{(i)} \mid y) \prod_{j=0}^{d-1} p(x_j^{(i)} \mid y)\right)
\end{align*}
where the 3 groups of learnable parameters, collectively denoted with $\vec{\Theta}$, are: 
\begin{enumerate}
    \item $p(y)$:
    \begin{itemize}
        \item prior probability to observe class $y$;
        \item there is one for each of the $k$ classes;
        \item it satisfies $\sum_{y=0}^{k-1} p(y)=1$.
    \end{itemize}
    \item $p(a \mid y)$:
    \begin{itemize}
        \item probability to observe an activity $a$ given the class $y$:
              \begin{align*}
                  p(a \mid y)=\Phi(a+1;\mu_y, \sigma_y)-\Phi(a;\mu_y, \sigma_y) \\
                  \mathrm{where} \: \: \Phi:= \mathrm{CDF(lognormal(\mu, \sigma))}
              \end{align*}
        \item $(\mu_y, \sigma_y)$ are the learnable parameters;
        \item there is one tuple for each of the $k$ classes.
    \end{itemize}
    \item $p(j \mid y)$: 
    \begin{itemize}
        \item probability to observe an activation over subreddit $j$ given the class $y$;
        \item the probability to observe $x$ independent activations in the subreddit $j$ given the class $y$ is \[
        p(x_j \mid y)=p_j(j \mid y)^{x_j}
        \] 
        \item there is one for each class ($k$) and each subreddit ($d$)
        \item it satisfies $\sum_{j=0}^{d-1} p(j \mid y)=1$.
    \end{itemize}
\end{enumerate}

\spara{Bayesian Models.}  
This formulation leads to three main models considered throughout the paper:  

\begin{enumerate}
    \item Standard Multinomial Naive Bayes (\texttt{NB}):  
    This model is trained using closed-form counting formulas to estimate the probabilities.

    \item Log-Normal Naive Bayes (\texttt{LogN NB}):
    In this model, the parameters of the log-normal (mean and standard deviation) are fitted to the data to model \(p(a \mid y)\).
    
    \item Semi-Supervised Naive Bayes (\texttt{SS NB}):
    This model uses the Expectation-Maximization algorithm to incorporate both labeled and unlabeled data during training.
    
    \item Semi-Supervised Log-Normal Naive Bayes (\texttt{SS LogN NB}):
    This hybrid model combines the semi-supervised approach with the log-normal distribution, fitting the activity (\(p(a \mid y)\)) while leveraging the EM algorithm.
\end{enumerate}
Notice that both \texttt{NB} and \texttt{LogN NB} are trained in a fully supervised manner, assuming all class labels (\(y\)) are given.

\spara{Semi-Supervised Learning.}  
In semi-supervised learning, the presence of unlabeled data introduces latent variables, making the problem more complex.
To address this issue, parameter estimation relies on the Expectation-Maximization (EM) algorithm, which iteratively refines the parameters by alternating between the following two steps:

\textit{E-step.} We define the conditional class probabilities given the current parameter $\vec{\Theta}^{t}$. For each sample $i$, and each label $y$ the probability is
\begin{align*}
    p(y \mid \vec{x}^{(i)};\vec{\Theta}^{t})= 
    \frac{p^{t}(y) p^{t}(a^{(i)} \mid y) \prod_{j=0}^{d-1} p^{t}_j(x_j^{(i)} \mid y)}{\sum_{y=0}^{k-1} p^{t}(y) p^{t}(a^{(i)} \mid y) \prod_{j=0}^{d-1} p^{t}_j(x_j^{(i)} \mid y)}
\end{align*} where $\sum_y p(y \mid \vec{x}^{(i)};\vec{\Theta}^{t}) = 1$. In the supervised setting, when $y^{(i)}$ is not missing ($y^{(i)}\neq -1$), there is only one term with probability one: $ p(y \mid \vec{x}^{(i)};\vec{\Theta}^{t})=1$ if $y=y^{(i)}, 0$ otherwise.

\textit{M-step.} By MLE, the three sets of parameters are updated in the following way.
\begin{enumerate}
    \item Class probability: $\forall y \in \{0, \ldots , k-1 \}$ 
    \[p^{t+1}(y)=\frac{\alpha_1+\sum_{i=1}^n p(y \mid \vec{x}^{(i)};\vec{\Theta}^{t})}{\alpha_1 k+ n}\]
    where $\alpha_1$ is the regularization term;
    \item conditional activity: $\forall y \in \{0, \ldots , k-1 \}$
    \begin{align*}
         \mu_y=\frac{1}{n}\sum_{i=1}^{n}\log{a^{(i)}} \\
         \sigma_y=\sqrt{\frac{1}{n}\sum_{i=1}^{n} \left( \log{a^{(i)}}- \mu_y \right)^2} 
    \end{align*}
    \item conditional activations: $\forall y \in \{0, \ldots , k-1 \} $ and $\forall j \in \{1, \ldots , d \} $
    \[p^{t+1}(j \mid y)=\frac{\alpha_2+\sum_{i=1}^n x_j^{(i)}p(y \mid \vec{x}^{(i)};\vec{\Theta}^{t})}{\alpha_2 d+ \sum_{j=0}^{d-1} \sum_{i=1}^n x_j^{(i)} p(y \mid \vec{x}^{(i)};\vec{\Theta}^{t})}\]
     where $\alpha_2$ is the regularization term.
\end{enumerate}

\newpage

\subsection{Full Experimental Setup}
We consider two main quantitative prediction tasks: classification and quantification of binary sociodemographic attributes:
\emph{(i) Year of Birth}: whether a user's year of birth is lower than the median of the user base;
\emph{(ii) Gender}: gender, considering only binary categories (Male and Female);
\emph{(iii) Partisan Affiliation}: party alignment with the U.S. two main parties (Democrat and Republican).

In contrast to~\citet{waller2021quantifying}, we mapped the self-declared age to the year of birth for each user.
This transformation ensures that the attribute remains static, and addresses the evolving nature of age over a time span longer than one year, as in our case study.

\label{sec:setup}
\spara{Benchmark and Parameters.}
We evaluate the following classifiers:
\emph{(i)} \texttt{Majority Model} (predicts the most frequent class in the dataset. This serves as a reference point for the minimum acceptable performance); 
\emph{(ii)} \texttt{WA Model} \cite{waller2021quantifying}; 
\emph{(iii)} \texttt{Random Forest (RF)} (with 50 estimators and max depth 10);
\emph{(iv)} \texttt{Multinomial Naive Bayes (NB)} (with additive smoothing parameter $\alpha=1$).

For quantification tasks, we apply the following quantification \emph{wrappers} on each classifier above:
\begin{itemize}

    \item \texttt{Classify \& Count (CC)}: Classifies each data point and returns the raw cardinality of each predicted class. This serves as a simple baseline.
    \item \texttt{Adjusted Classify \& Count (ACC)}: Adjusts the raw class proportions by fitting conditional distributions to correct potential biases in the CC method. This implementation follows the method proposed by \citet{moreo2021quapy}.
\end{itemize}

\paragraph{Extensions to Naive Bayes for Quantification.}
To evaluate common practices to enhance the quantification capabilities of the Naive Bayes model, we evaluate the following advanced variants:
\emph{(i)} \texttt{Semi-Supervised Naive Bayes (SS NB)};
\emph{(ii)} \texttt{Log-Normal Naive Bayes (logN NB)}, which models user activity as class-dependent);
\emph{(iii)} \texttt{Semi-Supervised Log-Normal Naive Bayes (SS logN NB)}, a hybrid approach.


\spara{Training Settings.}
While the method by \citet{waller2021quantifying} is unsupervised, the others require a training phase.
We compare two distinct training data settings.

\emph{True Supervision:} We use labels derived directly from explicit user self-declarations on Reddit.
These labels provide a reliable ground truth based on self-declared sociodemographic attributes and serve as a benchmark for evaluating model performance under ideal conditions.

\emph{Distant Supervision:}
We adopt a procedure inspired by~\citet{waller2021quantifying}:
training labels are based on their participation in discriminative subreddits associated with each class of each attribute.
For each attribute (\emph{year}, \emph{gender}, \emph{partisan}), we define two sets of five subreddits—one set corresponding to each label (e.g., \emph{young/old}, \emph{male/female}, \emph{democrat/republican}).
A user is labeled according to a specific class if their participation in one set of subreddits exceeds their participation in the opposite set for that attribute.
To ensure robust and unambiguous labeling, a user is included in the training set only if their participation difference between the two sets exceeds a threshold of three interactions.
This threshold minimizes noise and reduces the likelihood of mislabeling due to ambiguity.

The key distinction between the two data settings lies in their respective strengths and limitations.
Self-declared labels act as direct proxies for users’ sociodemographics, offering high-quality and explicit information.
However, these labels are often associated with a narrower range of subreddit activity, which limits the diversity of user behaviors represented in the training data, as shown in Figures~\ref{fig:regex1}, \ref{fig:regex3}.
Conversely, distant supervision broadens the scope of user activity by covering a wider range of users whose activity is orthogonal to the choice to be labeled.
This expands the dataset and captures more heterogeneous patterns of activity.
Nevertheless, distant supervision relies heavily on assumptions about the associations between subreddits and sociodemographic attributes, which may introduce noise and reduce labeling accuracy.

The class distribution for the classification task is depicted in Figure~\ref{fig:class_distr_true}.
To address class imbalances, we apply a Random Over Sampler, which ensures a balanced representation of both classes during training.
Furthermore, for training with distant supervision, we apply distant labeling to the entire set of users, regardless of the disclosed attributes, thereby enlarging the user base.

\spara{Evaluation Metrics.}
All metrics are computed using self-declarations only (also in the distant-supervision case).
For classification, performance is measured via average ROC AUC and F1-score, calculated over $100$ bootstrapped test samples, each comprising $20\%$ of the data.
For quantification, we use the Mean Absolute Error (MAE).
We divide the dataset into $70\%$ for training the models and $30\%$ held out as unseen data for building test sets, using $50$ different random seeds.
For each seed, we apply the Natural Prevalence Protocol (NPP)~\cite{moreo2021quapy} to generate $50$ test sets, each with class prevalences approximating the natural prevalence of the two classes for each attribute.
We compute Absolute Errors for each test set across all repetitions and seeds, then average across all seeds and repetitions to obtain the MAE.

\section{Dataset}
The ground truth for sociodemographic attributes is derived from user self-declarations collected across the \num{10000} most active subreddits between 2016 and 2020. After filtering out bots and non-coherent users, we retain approximately \num{400000} labeled instances each for Gender and Year of Birth, and roughly \num{4000} for Partisan Affiliation—highlighting the sharp class imbalance across attributes (Table~\ref{table:stats}).

\begin{figure*}[h!]
    \centering
    \begin{minipage}[t]{0.32\textwidth}
        \centering
        \includegraphics[width=\textwidth]{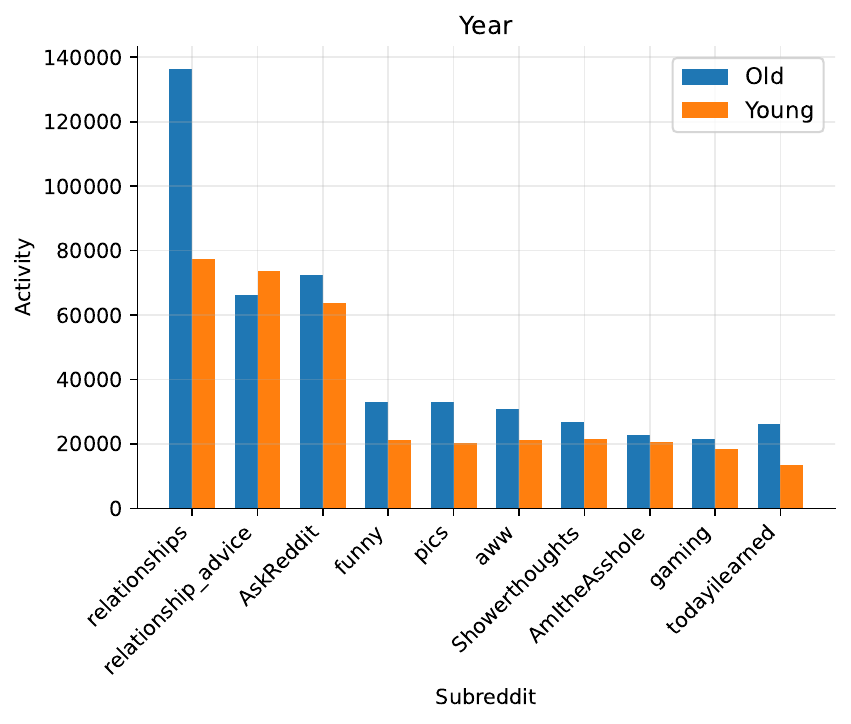}
        \textbf{(a)} Year of Birth
        \label{fig:regex1}
    \end{minipage}
    \hfill
    \begin{minipage}[t]{0.32\textwidth}
        \centering
        \includegraphics[width=\textwidth]{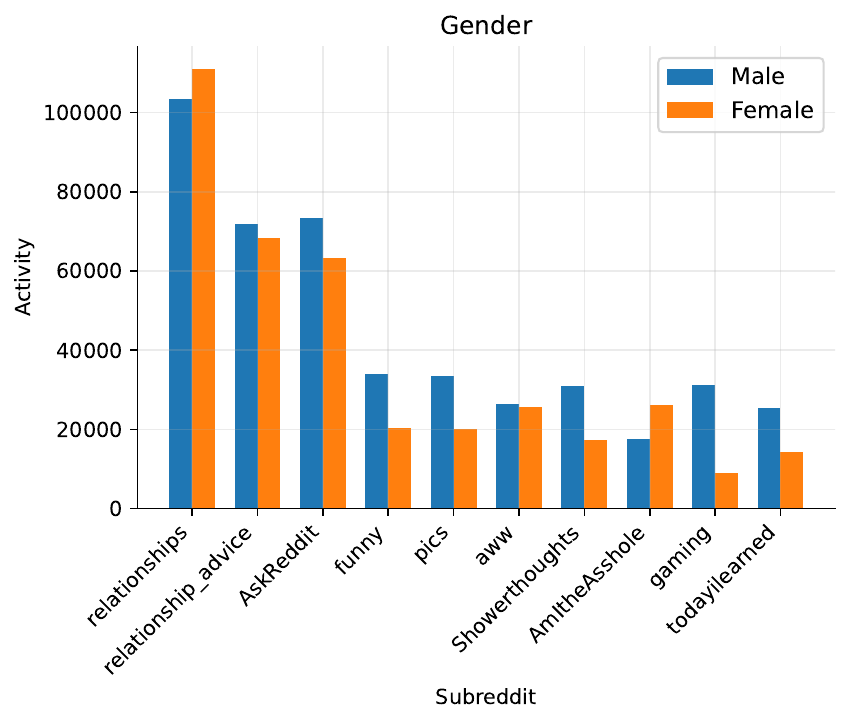}
        \textbf{(b)} Gender
        \label{fig:regex2}
    \end{minipage}
    \hfill
    \begin{minipage}[t]{0.32\textwidth}
        \centering
        \includegraphics[width=\textwidth]{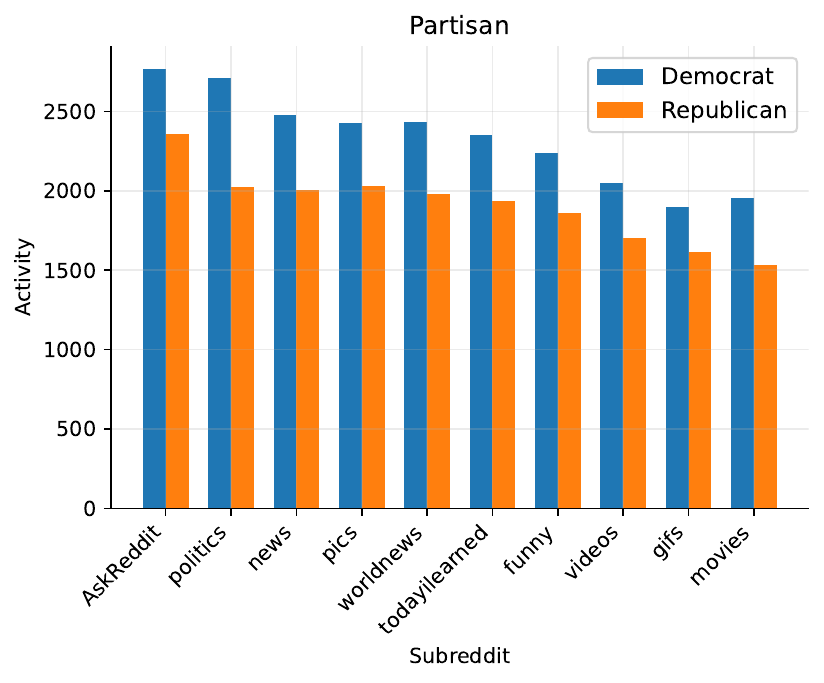}
        \textbf{(c)} Partisan Affiliation
        \label{fig:regex3}
    \end{minipage}

    \caption{Subreddit titles for socio-demographic disclosure: year of birth, gender, and partisan affiliation.}
    \label{fig:activity_distribution_combined}
\end{figure*}

Disclosure patterns vary across subreddits, as shown in Figures~\ref{fig:activity_distribution_combined}. For Year of Birth, subreddits such as \texttt{relationships} are highly discriminative, while gaming-related communities like \texttt{gaming} are more prominent for Gender. In contrast, Partisan Affiliation is signaled by political subreddits such as \texttt{politics} and \texttt{news}, reflecting the focus of those communities.

\begin{figure}
\includegraphics[width=.94\columnwidth, trim=0mm 6mm 0mm 7mm, clip]{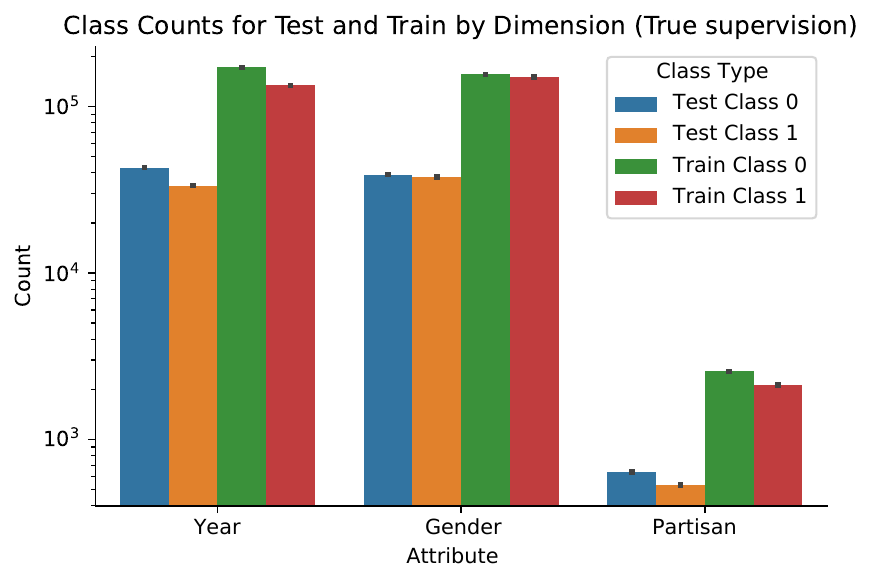}
\caption{(True Supervision) Distribution of the user-declared labels on the three attributes, divided into train and test set (80/20) for the classification task. Classes 0: Old, Male, Democrat. Classes 1: Young, Female, Republican.}
\label{fig:class_distr_true}
\end{figure}

Figure~\ref{fig:class_distr_true} shows the distribution of user-declared labels across the three attributes, using an 80/20 train-test split for classification. The number of declarations for Partisan Affiliation is significantly lower—roughly an order of magnitude smaller—than for Year and Gender.

\begin{figure}[t]
    \centering
    \includegraphics[width=0.4\textwidth]{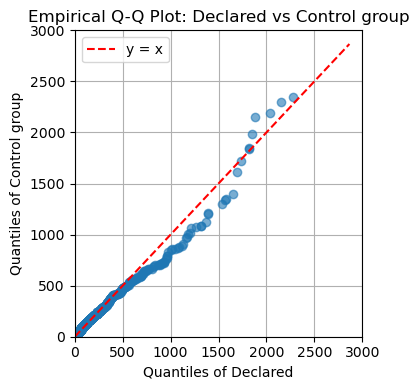}
    \caption{Empirical Q-Q plot comparing subreddit participation counts for declared vs. activity-matched control users. Each point represents a subreddit. The strong diagonal alignment suggests similar community engagement across both groups.}
    \label{fig:qq_plot}
\end{figure}

To evaluate whether self-declaring users systematically differ from the broader Reddit population, we compare aggregate subreddit participation patterns between declared users and a matched control group. For each declared user, we randomly sample a non-declared user with similar overall activity (i.e., total number of comments), and compute the total number of comments per subreddit for both groups. Figure~\ref{fig:qq_plot} shows a Q-Q plot where each point represents a subreddit, with the x-axis indicating its total participation count among declared users and the y-axis the corresponding count among control users. We exclude five subreddits with more than \num{3000} comments to reduce the influence of long-tail popularity effects. The near-diagonal alignment suggests that declared users engage with subreddits in broadly similar proportions to activity-matched users, indicating no major structural bias in subreddit preferences.

\begin{figure*}[h!]
    \centering
    \begin{minipage}[t]{0.32\textwidth}
        \centering
        \includegraphics[width=\textwidth]{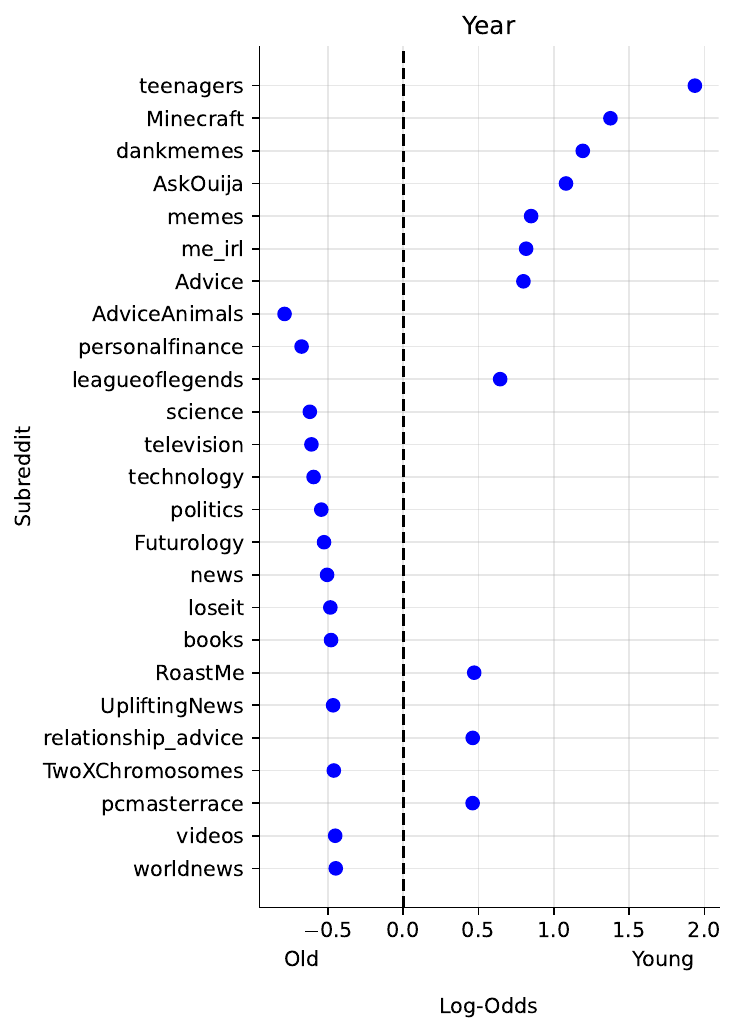}
        \textbf{(a)} Year of Birth
        \label{fig:importance-year}
    \end{minipage}
    \hfill
    \begin{minipage}[t]{0.32\textwidth}
        \centering
        \includegraphics[width=\textwidth]{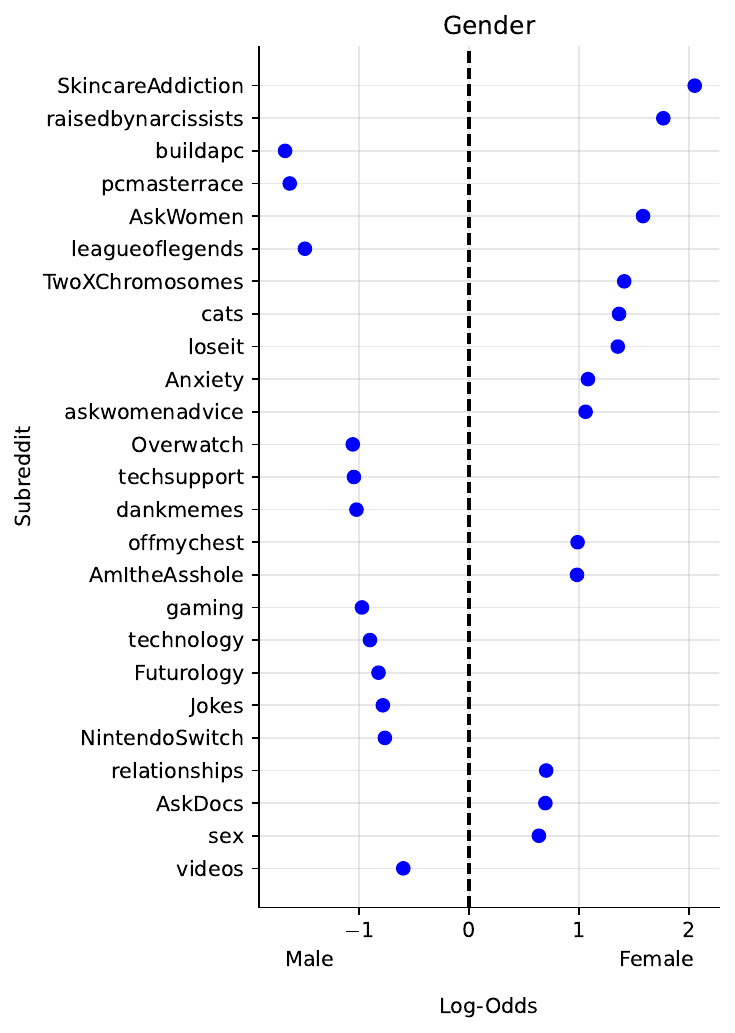}
        \textbf{(b)} Gender
        \label{fig:importance-gender}
    \end{minipage}
    \hfill
    \begin{minipage}[t]{0.32\textwidth}
        \centering
        \includegraphics[width=\textwidth]{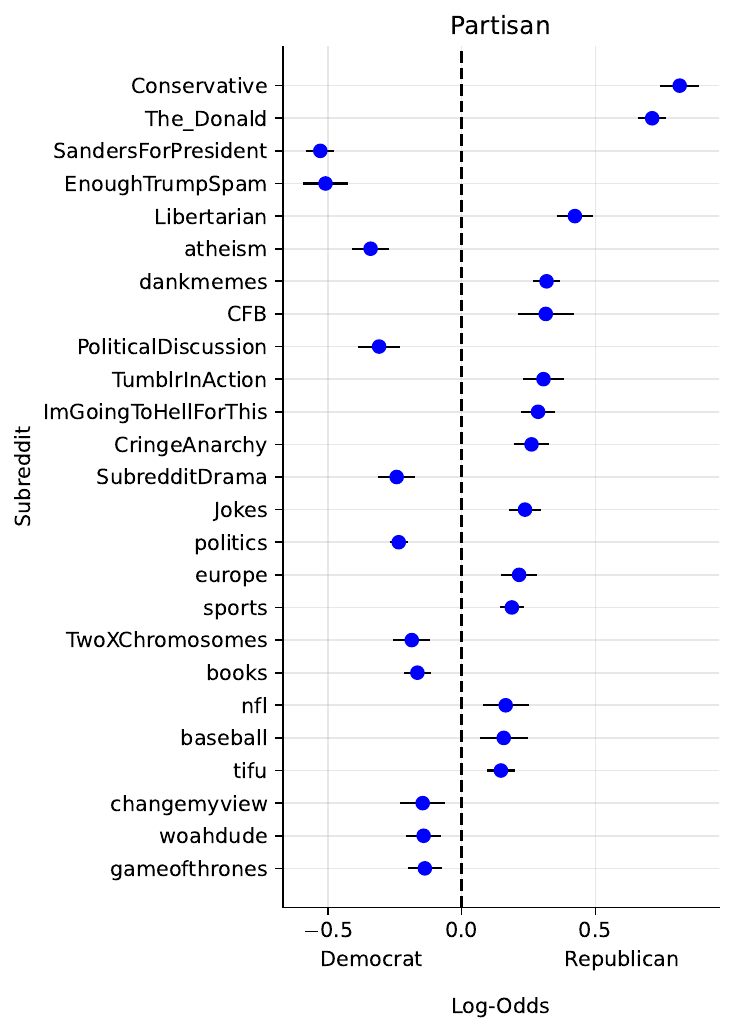}
        \textbf{(c)} Partisan Affiliation
        \label{fig:importance-partisan}
    \end{minipage}

    \caption{Top 25 most predictive subreddits for each classification task, selected among the top 1\% most active subreddits.}
    \label{fig:importance_combined}
\end{figure*}

\begin{figure*}[b]
\includegraphics[width=.94\textwidth]{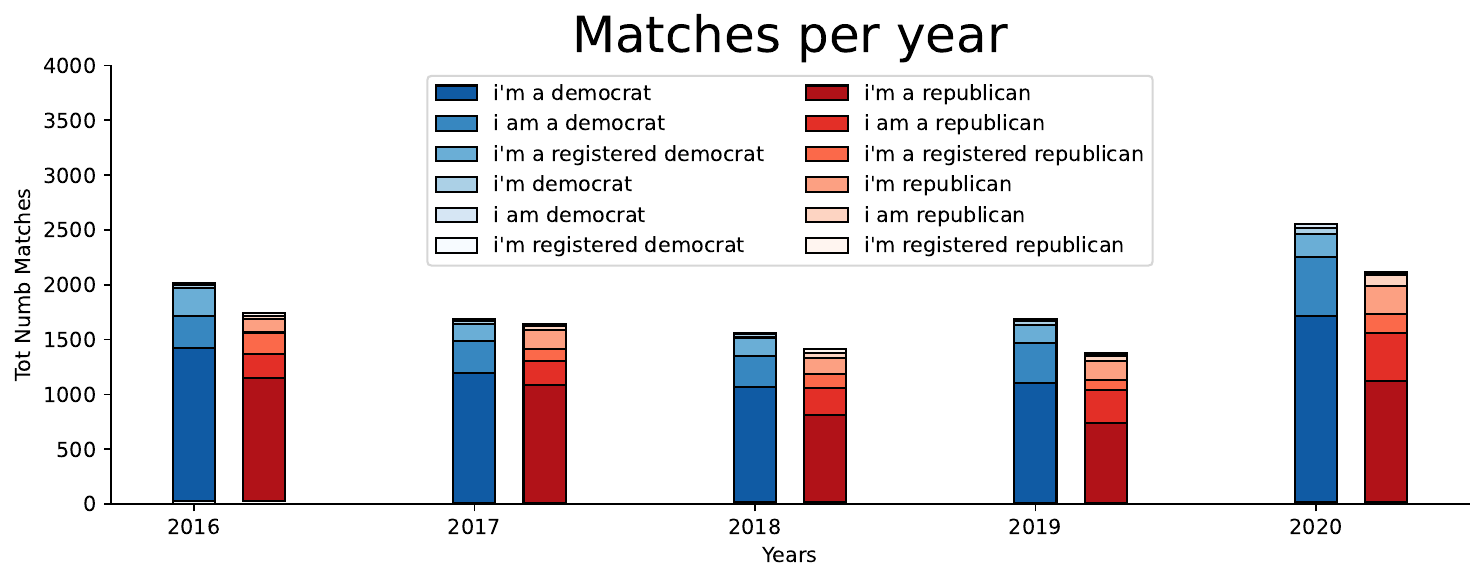}
\caption{Distribution of matched regular expressions for partisan affiliation per year.}
\label{fig:regex-pol}
\end{figure*}

\end{document}